\documentclass{article}
\usepackage{arxiv}
\usepackage[utf8]{inputenc}
\usepackage[T1]{fontenc}
\usepackage{xcolor}
\usepackage[sort&compress, round]{natbib}
\definecolor{blendedblue}{rgb}{0.2, 0.2, 0.6}
\usepackage{graphicx}
\usepackage{enumerate}
\usepackage[bookmarks   = true,
            citecolor   = blendedblue,
            colorlinks  = true,
            linkcolor   = blendedblue,
            urlcolor    = blendedblue,
            citecolor   = blendedblue,
            linktocpage = false,
            hyperindex  = true]{hyperref}
\usepackage{booktabs}
\usepackage{amsfonts}
\usepackage{nicefrac}
\usepackage{amsmath, amssymb, amsthm}
\usepackage{url}
\usepackage{doi}
\usepackage{bm}
\usepackage{todonotes}
\usepackage{fontawesome}
\usepackage{mathtools}
\usepackage{lmodern}
\usepackage{siunitx}
\usepackage{booktabs}
\usepackage{etoolbox}
\usepackage{calc}
\usepackage{enumitem}
\usepackage{dsfont}
\usepackage{multirow}
\usepackage{array}
\usepackage{graphicx}
\usepackage{tikz}
\definecolor{blendedblue}{rgb}{0.2, 0.2, 0.6}

\usepackage{enumerate} 
\usepackage{orcidlink}
\usepackage[labelsep=period]{caption}

\renewcommand{\headeright}{}

\usepackage{parskip}
\makeatletter 
\renewcommand\@biblabel[1]{#1.} 
\makeatother

\title{Confounder-adjusted Covariances of System Outputs and Applications to Structural Health Monitoring}
\renewcommand{\shorttitle}{Confounder-adjusted Covariances of System Outputs}
\date{}

\author{Lizzie Neumann\orcidlink{0000-0003-2256-1127}\\
 	Dept. of Mathematics and Statistics\\
	School of Economics and Social Sciences\\
        Helmut Schmidt University\\
	Hamburg, Germany\\
	\texttt{neumannl@hsu-hh.de} \\
	\And
        Philipp Wittenberg\orcidlink{0000-0001-7151-8243}\\
 	Dept. of Mathematics and Statistics\\
	School of Economics and Social Sciences\\
        Helmut Schmidt University\\
	Hamburg, Germany\\
	\texttt{pwitten@hsu-hh.de} \\
	\And
        Alexander Mendler\orcidlink{0000-0002-7492-6194}\\
        Dept. of Materials Engineering\\
        TUM School of Engineering and Design\\
	Technical University of Munich\\
        Munich, Germany\\
	\texttt{alexander.mendler@tum.de}\\
	\And 
        Jan Gertheiss\orcidlink{0000-0001-6777-4746}\\
 	Dept. of Mathematics and Statistics\\
 	School of Economics and Social Sciences\\
        Helmut Schmidt University\\
	Hamburg, Germany\\
	\texttt{gertheij@hsu-hh.de} \\
}

\newcommand\copyrighttext{%
  \scriptsize This is the peer-reviewed version of the following article: Neumann L., Wittenberg, P., Mendler, A., Gertheiss, J. (2024). Confounder-adjusted Covariances of System Outputs and Applications to Structural Health Monitoring. Accepted for publication in \textit{Mechanical Systems and Signal Processing} © 2024. This manuscript version is made available under the \href{https://creativecommons.org/licenses/by-nc-nd/4.0/}{CC-BY-NC-ND 4.0} license.}
\newcommand\copyrightnotice{%
\begin{tikzpicture}[remember picture,overlay]
\node[anchor=south,yshift=10pt] at (current page.south) {\fbox{\parbox{\dimexpr\textwidth-\fboxsep-\fboxrule\relax}{\copyrighttext}}};
\end{tikzpicture}%
}

\begin{document}	
\copyrightnotice
\maketitle

\begin{abstract}
Automated damage detection is an integral component of each structural health monitoring (SHM) system. Typically, measurements from various sensors are collected and reduced to damage-sensitive features, and diagnostic values are generated by statistically evaluating the features. Since changes in data do not only result from damage, it is necessary to determine the confounding factors (environmental or operational variables) and to remove their effects from the measurements or features. Many existing methods for correcting confounding effects are based on different types of mean regression. This neglects potential changes in higher-order statistical moments, but in particular, the output covariances are essential for generating reliable diagnostics for damage detection. This article presents an approach to explicitly quantify the changes in the covariance, using conditional covariance matrices based on a non-parametric, kernel-based estimator. The method is applied to the Munich Test Bridge and the KW51 Railway Bridge in Leuven, covering both raw sensor measurements (acceleration, strain, inclination) and extracted damage-sensitive features (natural frequencies). The results show that covariances between different vibration or inclination sensors can significantly change due to temperature changes, and the same is true for natural frequencies. To highlight the advantages, it is explained how conditional covariances can be combined with standard approaches for damage detection, such as the Mahalanobis distance and principal component analysis. As a result, more reliable diagnostic values can be generated with fewer false alarms.
\end{abstract}
\bigskip
\noindent%


\section{Introduction}
\label{neumann:sec_intro}
Structural health monitoring (SHM) describes the automated damage assessment of engineered systems in civil, mechanical, and aerospace engineering based on sensors \citep{Farrar.Worden_2013}. The diagnostic chain includes the acquisition of measurement data, the extracting of damage-sensitive features, and their statistical evaluation, e.g., based on covariance matrices. Typically, feature changes are evaluated for damage detection, but no changes in the covariances are assumed throughout training, validation, and testing. However, this article will demonstrate that this assumption is often unjustified in civil engineering. To remedy this, this paper develops an approach that quantifies the uncertainty of \textit{system outputs}, i.e., measurements or damage-sensitive features, for varying confounding variables (environmental or operational parameters (EOP)) in terms of conditional covariances. 

Civil engineering structures, such as bridges, are exposed to changing environmental and operational conditions, causing most system response quantities to fluctuate even in the absence of damage, e.g., 
due to changes in temperature, humidity, traffic, wind, and solar radiation. These dependencies are inevitable for static response measurements such as strain~\citep{Kromanis.Kripakaran_2016, Liang.etal_2009} and inclination~\citep{Han.etal_2021, Ju.etal_2023, Lee.etal_2019}, and dynamic ones (e.g., acceleration, natural frequencies). For instance, the relationship between temperature and natural frequencies is described by \citet{Han.etal_2021}, \citet{Moser.Moaveni_2011}, and \citet{Xia.etal_2006}, while \citet{Xia.etal_2006} and \citet{Zhang.etal_2012} focus on damping ratios.
In some cases, the changes in system response quantities due to environmental changes are more pronounced than those due to structural damage.
Therefore, methods for removing confounding effects are essential when analyzing the system output data. In SHM, this is also known as ``data normalization'' \citep{Farrar.Worden_2013}, and various comprehensive literature reviews are available \citep[e.g.,][]{Han.etal_2021, Wang.etal_2022}. 
Whereas some damage diagnosis methods inherently include data normalization in the training phase, e.g., neural networks, other approaches require data normalization as a separate step in the modular diagnostic chain, which typically happens after the extraction of damage-sensitive features. Afterward, the features are handed over to other diagnosis methods based on statistical process control (SPC), model updating approaches, or others. An alternative categorization is based on whether the confounder variables are measured or not: 
%
\textit{Supervised methods} describe a group of algorithms that require the confounder variables to be measured in combination with the system outputs.
A popular approach is regressing the sensor measurements of interest or derived features such as natural frequencies on the confounders because it leads to multi-dimensional plots that users can easily interpret, so-called ``response surface models'' \citep{Worden.etal_2016, Worden.Cross_2018}. After model training, the predicted values can, for example, be subtracted from the observed response values, which is known as the ``subtraction method'' \citep{Worden.etal_2016, Worden.Cross_2018}. 
In biostatistics (where measurements typically have to be adjusted for age, sex, etc.), this process is also called ``residualization'' \citep{Zhou.Wright_2015, Donovan.etal_2023} because the residuals from the confounder model are used for further analysis. In the context of SHM, \citet{Maes.etal_2022}, for instance, used linear regression and the resulting normalized misfits for monitoring. They showed the limitations of this approach due to the linearity assumption of the system output-confounder relationship.
\citet{Moser.Moaveni_2011} presented various methods and concluded that non-linear, fourth-order polynomial regression works best for temperature compensation. They also presented a `bilinear' approach to model the kink around zero degrees (Celcius) in the function that relates system response quantities to ambient temperature records. 
\citet{Worden.Cross_2018} used a Bayesian treed Gaussian process model, which is another method that can handle (potentially) non-linear environmental influences on system outputs. 
\citet{Wang.etal_2022} give an overview of different possibilities to analyze and model the association between system outputs and temperature, including a (non-)linear autoregressive model with exogenous input (ARX), neural networks, and further regression methods. 
\citet{Roberts.et.al_2024} utilized multivariate nonlinear regression along with an updating scheme to account for EOPs. Initially, nonlinear stepwise regression was employed to update the univariate nonlinear regression models until no further improvements could be made, and only the influential EOPs were selected. Subsequently, the results were used in the multivariate nonlinear regression model.
\citet{Xu.et.al_2023} examined the relationship between modal frequencies and environmental factors, such as temperature and relative humidity, through correlation analysis. They found that temperature has the most significant impact. To address this issue, they employed a support vector regression model to remove the influence of temperature on modal frequencies.
%
\textit{Unsupervised methods}, on the other hand, describe an alternative group of methods that does not require the confounder to be measured to remove its effect on the system outputs.
Representative methods include principal component analysis (PCA), intelligent feature selection (which are robust to environmental changes), and various other machine learning-based methods. 
For example, \citet{Reynders.etal_2014} extended standard, linear PCA to a kernel-based PCA that allows for a non-linear environmental model. \citet{Maes.etal_2022} used a robust version of PCA and compared it to common PCA and the linear regression method mentioned above. They conclude that robust PCA represents the behavior of the data well, however, they also point out that the differences to the standard PCA are small, at least for the analyzed set of data. 
\citet{Entezami.etal_2020} proposed an unsupervised learning method based on time series analysis, deep learning, and the Mahalanobis distance for feature extraction, dimensional reduction, and classification, respectively. \citet{Wang.etal_2022} provide an overview of different unsupervised methods and classify them as feature extraction, cointegration, and sequence decomposition-based methods, where the feature extraction-based methods consist mainly of different versions of PCA and neural network methods. 
\citet{Cross.etal_2011} use cointegration to linearly combine response variables to create stationary residuals and use these for monitoring. 

This paper focuses on supervised methods, particularly response surface modeling, where both the system outputs of interest and the potential confounder(s) are available. As pointed out above, great progress has been made over the last years in this area, particularly in terms of multivariate, nonlinear modeling. However, all existing methods for response surface modeling mentioned above focus on mean response values given the confounder(s) and do not normalize higher-order statistics. To emphasize this aspect, the approaches are categorized under the term \emph{mean regression} in statistics. Experience with large engineering systems, however, has shown that uncertainty quantification (the estimation and analysis of the measurement errors and variability) is equally important as evaluating the mean values, and thus, monitoring systems often use higher-order statistical moments such as covariances for uncertainty quantification and damage detection \citep{Doehler.etal_2014, Montgomery_2007}. 
The problem with mean regression and the subtraction method is that it may miss effects on higher-order moments, such as covariances. Consequently, if a damage identification tool only uses mean regression and residualization to account for environmental effects but covariances of system outputs at some later point along the diagnostic chain, an alarm might be issued due to changes in the covariance caused by the confounder variable.
In other words, this procedure may lead to many false alarms, and the structure might be closed for no reason. Therefore, an alternative approach that adequately accounts for the confounding effects on the covariance is needed to reduce the number of false alarms.

However, the literature on the environmental effects on covariances of system outputs is very limited, particularly in the context of SHM. 
\citet{Neumann.Gertheiss_2022} estimated the partial covariance by regressing the data (in a non-linear way) on the temperature data using penalized regression splines and then calculated the covariance of the residuals. 
\citet{viefhues2020fault,Viefhues2022} developed a diagnostic test, similar to the Mahalanobis distance, that considers different covariances in the training and testing data. 
Furthermore, \citet{Viefhues.etal_2021} did a laboratory test in which they measured the acceleration on a concrete beam for five different temperatures in a climate chamber. Then, they estimated the (conditional) covariance for these five temperature values and interpolated the covariances for temperature values in between.
This article considers a different situation, where the temperature has to be treated as a continuous variable with values scattered over a wide temperature range. The main contribution of this paper is to show how conditional covariances of system outputs can be estimated in this case using a kernel-based smoothing operator. 
In addition to changes in the mean value due to environmental effects, the proposed method accounts for changes in the (co-)variance, which significantly improves the reliability of the damage detection algorithms.

The remainder of this article is structured as follows. Section~\ref{neumann:sec_method} discusses the concepts of partial and conditional covariances and proposes a non-parametric estimate of the conditional covariance. Section~\ref{neumann:sec_shm} explains how the developed methods can be combined with well-established methods, such as the Mahalanobis distance and PCA, to obtain damage-sensitive features and diagnostic tests that are robust to environmental changes. Section~\ref{neumann:sec_validation_of_ana_method} presents a Monte Carlo simulation study to validate the non-parametric estimate of the conditional covariance. Section~\ref{neumann:sec_application_SHM_data} illustrates the application of the proposed methodology on two SHM data sets and several sensor types. Finally, we provide some concluding remarks and an outlook in Section \ref{neumann:sec_conclusion}. 

\section{Partial and Conditional Covariances}
\label{neumann:sec_method}
This section discusses two approaches to obtain confounder-adjusted covariances between system outputs. In Section~\ref{neumann:sec_partial_covariance_response_surface}, we will shortly revisit the concept of response surface modeling and the subtraction method, whereas in Section~\ref{neumann:sec_nonoparm_est_cond_cov}, a novel method is proposed that can estimate conditional covariances. A summary of important symbols and the nomenclature that we will use throughout the paper is given in Table~\ref{neumann:tab_nomenclature}. The methods presented can be directly applied to sensor measurements (e.g., strains, inclinations, accelerations) or extracted damage-sensitive features (e.g., eigenfrequencies). That is why the more general term `system output(s)' will be used in this paper.

\begin{table}[h]
    \centering
    \footnotesize
    \caption{Summary of important symbols and nomenclature used in this paper.}
    \label{neumann:tab_nomenclature}
    \begin{tabular}{l l}
    \toprule
        $\mathbf{x}$ & vector of system outputs $x_1,\ldots,x_p$ (measurements or extracted features) \\
        $z$ & confounder (e.g., temperature) \\
        $\textbf{m}(z)$ & conditional mean vector of $\mathbf{x}$ for a given $z$\\
        $\boldsymbol{\Sigma}(z)$ & conditional covariance matrix of $\mathbf{x}$ for a given $z$\\
        $\sigma_{j,k}(z)$ & conditional covariance of system outputs $x_j$ and $x_k$ for a given $z$, which equals entry $(\boldsymbol{\Sigma}(z))_{j,k}$\\
        $K_h(z)$ & kernel function with bandwidth $h$\\
        $Q_{m_k}(h)$ & (quadratic) loss with respect to the mean of system output $x_k$ as a function of $h$\\
        $Q_{\sigma_{j,k}}(h)$ & (quadratic) loss with respect to the covariance of system outputs $x_j$ and $x_k$ as a function of $h$ \\
        $Q_{\boldsymbol{\Sigma}}(h)$ & (quadratic) loss with respect to the entire covariance matrix of system outputs as a function of $h$ \\
        $\boldsymbol{\Lambda}(z)$ & (diagonal) matrix of (conditional) eigenvalues $\lambda_1(z),\ldots,\lambda_p(z)$ of $\boldsymbol{\Sigma}(z)$\\
        \textbf{A}$(z)$ & matrix of (conditional) eigenvectors/principal components $\mathbf{a}_1(z),\ldots,\mathbf{a}_p(z)$ of $\boldsymbol{\Sigma}(z)$\\
        $\mathbf{s}_i$ & vector of principal component scores for instance $i$ \\
        \bottomrule
    \end{tabular}
\end{table}

\subsection{Response Surface Modeling and Partial Covariances}
\label{neumann:sec_partial_covariance_response_surface}
Let $\mathbf{x} = (x_1,\dots, x_p)^\top\in\mathbb{R}^{p}$ be a $p$-dimensional random (output) vector and $z\in\mathbb{R}$ a potential confounder. Then, the standard approach of response surface modeling uses a regression function $\mathbf{f}_\mathbf{x}(z)$ to account for the association between system outputs $\mathbf{x}$ and the covariate $z$ in terms of

\begin{equation*}
    \mathbf{x} = \mathbf{f}_\mathbf{x}(z) + \mathbf{\bm{\varepsilon}},
\end{equation*}

where $\mathbf{\bm{\varepsilon}} = (\varepsilon_1, \dots, \varepsilon_p)^\top$ is an error term and $\mathbf{f}_\mathbf{x}(z) = (f_{1}(z), \dots, f_{p}(z))^\top$. Then, further analyses would use the residuals, i.e., estimates of $\mathbf{\bm{\varepsilon}}$. 
As the introduction mentions, this procedure is called ``residualization'' in biostatistics or the ``subtraction method'' in the SHM literature. The covariance of the residuals is called \emph{partial} covariance.

Various methods and algorithms can be used to estimate the (potentially) non-linear regression functions $f_k(z)$, $k=1,\ldots,p$. \citet{Worden.Cross_2018}, for instance, used Gaussian process (GP) regression, while \citet{Neumann.Gertheiss_2022} used penalized regression splines, see also Section~\ref{neumann:sec_intro}.
However, \citet{Donovan.etal_2023} demonstrated that this approach may miss confounding effects on the covariance of $\mathbf{x}$, which is particularly harmful if the monitoring scheme uses these covariances to detect potential damages.

\subsection{Conditional Covariances}
\label{neumann:sec_nonoparm_est_cond_cov}
As an alternative to residualization and partial covariances, this section introduces a non-parametric, Nadaraya-Watson-type approach \citep{Nadaraya_1964, Watson_1964} to estimate the \emph{conditional} covariance matrix as a---potentially nonlinear---function of the confounder $z$.)
In that case, the estimator of the conditional covariance matrix $\boldsymbol{\Sigma}(z)$ of $\mathbf{x}$ given a confounder $z$ has the form \citep{Yin.etal_2010} 
\begin{equation}
    \label{neumann:eq_Sdef}
	\hat{\boldsymbol{\Sigma}}(z; h) = \left\{\sum_{i=1}^n K_h(z_i - z)\left[\mathbf{x}_i - \hat{\mathbf{m}}(z_i)\right]\left[\mathbf{x}_i - \hat{\mathbf{m}}(z_i)\right]^\top\right\}\left\{\sum_{i=1}^n K_h(z_i - z)\right\}^{-1},
\end{equation}
where $\mathbf{x}_i = (x_{i1}, \dots, x_{ip})^\top$, $i=1,\ldots,n$, are observations of $\mathbf{x}$, and $z_i$ is the associated confounder variable (e.g., temperature). 
The term $K_h(\cdot)$ is a kernel function with bandwidth $h$, and $\hat{\mathbf{m}}(z_i)$ is an estimate of the mean of $\mathbf{x}$ at $z_i$. For the mean, we can also use a kernel estimate 
\begin{equation}
    \hat{\mathbf{m}}(z; h) = \left\{\sum_{i=1}^n K_h(z_i - z)\mathbf{x}_i\right\}\left\{\sum_{i=1}^n K_h(z_i - z)\right\}^{-1}
    \label{neumann:eq_mdef}
\end{equation}
or any other method, such as penalized regression splines \citep{Eilers.Marx_1996, Neumann.Gertheiss_2022} or local polynomial regression \citep{Cleveland.etal_2017}. For the kernel functions $K_h$, we can use any symmetric probability density function $K(u)$ that is scaled by $K_h(u) = h^{-1}K(u/h)$, where $h > 0$ is the so-called bandwidth. This paper uses a Gaussian kernel, which equals a normal density with a zero mean value. The bandwidth $h$ is the smoothing parameter of the kernel. The higher $h$, the wider the kernel and the smoother the estimated covariance or mean function. For the covariance and mean estimation from Eq.~\eqref{neumann:eq_Sdef} and \eqref{neumann:eq_mdef}, the user can choose different bandwidth parameters. It is even possible to choose a different bandwidth parameter for each component of the conditional mean and conditional covariance \citep{Yin.etal_2010}, i.e., for each system output and each pair thereof. For example, the estimated (conditional) covariance of $x_j$ and $x_k$ is given by the matrix entry $(\hat{\boldsymbol{\Sigma}}(z; h))_{j,k} = \hat{\sigma}_{j,k}(z; h)$, where $j$ is the row and $k$ is the column index. 
Replacing the one global $h$ in Eq.~\eqref{neumann:eq_Sdef} with a specific $h_{j,k}$ for each covariance matrix entry allows for the adjustment of the smoothing parameter for different system outputs.
In summary, we obtain an estimate of the covariance matrix for any specific confounder value (within a reasonable interval), as the conditional covariance from Eq.~\eqref{neumann:eq_Sdef} is a function of the confounder $z$ (e.g., temperature). The function is continuous if the kernel function and the estimate of the (conditional) mean vector are continuous in $z$. Note that conditional and partial covariances are equivalent if the system outputs and the confounder jointly follow a (multivariate) normal distribution, but this assumption is not valid for most real-world applications. 

\subsection{Optimizing the Bandwidth}
\label{neumann:sec_bandwidth}
The developed approach requires one user-defined input parameter, the bandwidth $h$. The bandwidth influences the result, and the selection should be automated, particularly when an individual bandwidth parameter is used for each entry in the covariance matrix from Eq.~\eqref{neumann:eq_Sdef}. The suggested approach to finding the optimal bandwidth is to divide the reference data set into a training set and a validation set, to treat the bandwidth as a hyperparameter, and to define a loss function that penalizes large deviations between the reconstructed signal and the measured one. The optimal bandwidth for each component of the mean function $\mathbf{m}(z)$ can be found through the minimum of the following loss function
\begin{equation}
    \label{neumann:eq_Qmean}
    Q_{m_k}(h) = \sum_{i=1}^{\tilde{n}} [x_{ik} - \hat{m}_k(z_i;h)]^2,
\end{equation}
where $\hat{m}_k(z_i;h)$ is the $k$-th component of the estimated mean vector $\hat{\mathbf{m}}(z_i)$, $k = 1,\dots,p$, obtained from the training data for the bandwidth $h$.
Afterward, we can apply the same logic to obtain the smoothing parameter for the covariance estimate by minimizing
\begin{equation}\label{neumann:eq_QSig}
    Q_{\boldsymbol{\Sigma}}(h) = \sum_{j,k=1}^{p}\sum_{i=1}^{\tilde{n}} \left\{[x_{ij} - \hat{m}_j(z_i)][x_{ik} - \hat{m}_k(z_i)] - (\hat{\boldsymbol{\Sigma}}(z_i; h))_{j,k}\right\}^2 = \sum_{j,k=1}^{p} Q_{\sigma_{j,k}}(h) 
\end{equation}
using the validation data $i=1,\ldots,\tilde{n}$, where $(\hat{\boldsymbol{\Sigma}}(z_i; h))_{j,k}$ is the entry in row $j$ and column $k$ of the covariance matrix $\hat{\boldsymbol{\Sigma}}(z_i; h)$ obtained from training data with bandwidth $h$, cf. Eq.~\eqref{neumann:eq_Sdef}, and
$\hat{m}_k(z_i)$ is the $k$th component of $\hat{\mathbf{m}}(z_i)$ which resulted from Eq.~\eqref{neumann:eq_Qmean} (or the method of choice for mean regression). 
If we use specific bandwidths $h_{j,k}$ for the covariances in Eq.~\eqref{neumann:eq_Sdef}, we minimize the squared loss $Q_{\sigma_{j,k}}$ from Eq.~\eqref{neumann:eq_QSig} separately for each combination $(j,k)$. Alternatively, we can use $K$-fold cross-validation and split the data into several training and validation data sets. That means the data is split into $K$ parts, where one part is left out for validation, and $K-1$ parts are used for training. This is repeated $K$ times, and the results are averaged to obtain a one-dimensional quantity, analogously to Eq.~\eqref{neumann:eq_QSig}. 

The bandwidth parameter optimization scheme requires the reference data to be split into training and validation data, but the results depend on the user-defined data organization, and the following aspects should be considered:
Firstly, in long-term monitoring applications, reference data from several weeks, months, or even years is available and can be used for tuning hyperparameters such as the bandwidth. 
This data might exhibit temporal correlations, and to consider them during the bandwidth tuning, representative validation sample sequences of entire days or weeks should be drawn from the available data set. 
Secondly, confounder values should be equally represented in the training and validation set. We should not simply use, for example, the last few weeks of data available for validation, because the training and validation data should be comparable concerning the distribution of the confounder $z$. Therefore, we recommend defining several sub-intervals for the confounder $z$ and sampling the same proportion of data sequences from each interval, with the data sequences being assigned to the intervals according to their mean $z$-values.

\section{Confounder-adjusted Methods for SHM}
\label{neumann:sec_shm}
The conditional covariance matrix presented above can be used to adjust the Mahalanobis distance (MD) and principal component analysis (PCA) for confounding effects. MD and PCA use the covariance matrix of system outputs as an essential building block. They are often employed in SHM but, so far, disregard changes in the covariance due to confounding effects. In this section, modified versions are developed, based on the conditional covariance from Section~\ref{neumann:sec_nonoparm_est_cond_cov}, that lead to more reliable damage detection results and fewer false alarms if the covariance changes. 
\subsection{Conditional Mahalanobis Distance}
\label{neumann:sec_cond_md}
The squared Mahalanobis distance is a standard tool in statistical process control and SHM. In the univariate case (with a single measurement or feature), it quantifies the significance of changes in observed outputs through the number of standard deviations from a reference value squared. In general, it evaluates the deviation of the measured system outputs from a reference value (e.g., the mean value from training). In order to obtain a standardized version of these deviations, the covariance matrix of the system outputs is needed, and typically, a constant covariance matrix is assumed. In what follows, a modified version is proposed, that is, the conditional Mahalanobis distance (CMD). It is defined as
\begin{equation}
    d_\text{CMD}^2(z) = 
    (\mathbf{x} - \mathbf{m}(z))^\top 
    \boldsymbol{\Sigma}(z)^{-1} 
    (\mathbf{x} - \mathbf{m}(z)),
    \label{neumann:eq_mahalanobis}
\end{equation}
where $\mathbf{x} \in \mathbb{R}^{p}$ is a $p$-dimensional vector of observations,
$\mathbf{m}(z)\in \mathbb{R}^{p}$ is the conditional mean vector that can be estimated through Eq.~\eqref{neumann:eq_mdef}, and 
$\boldsymbol{\Sigma}(z)\in \mathbb{R}^{p\times p}$ is the conditional covariance matrix with the in Eq.~\eqref{neumann:eq_Sdef} proposed estimator, both of which are evaluated for the specific state of the confounder variable. 
The conditional Mahalanobis distance thus takes the value of the confounder $z$ 
into account through both the mean and the covariance function. Considering the conditional covariance means, among other things, that the diagnostic considers that system outputs may exhibit more variation for some temperature values than others.
An illustration will be provided in Section~\ref{neumann:sec_cond_mdKW51}, where the CMD is applied to the modal frequencies of a steel railway bridge.

\subsection{Conditional Principal Component Analysis and Feature Extraction}
\label{neumann:sec_pca}
In SHM, principal component analysis (PCA) is a universal mathematical tool for dimension reduction, anomaly detection, data normalization~\citep{Flexa.etal_2019, Reynders.etal_2014}, and feature extraction~\citep{Tibaduiza.etal_2016, Zhu.etal_2019}, where some PCA-based features are (supposedly) insensitive to environmental changes \citep{Kumar.etal_2020}. PCA can be understood as a rotation of the coordinate system that best aligns with the main directions of variation in the data. In the rotated coordinate system, the user can select some sources of variation and use only those  
for further data analysis and damage detection. It is assumed that the first principal components mainly account for operational or environmental effects (e.g., temperature effects), and the remaining components can either be used as damage-sensitive features \citep{Cross.etal_2012}, or they can be used to reconstruct the system outputs that are not affected by confounders.
Typically, PCA is an unsupervised method that only needs the output covariance matrix as input, but it does not consider confounder-induced changes in the covariance. 
Therefore, this section presents a novel version that explicitly accounts for confounder effects by applying the eigendecomposition to the conditional output covariance matrix from Eq.~\eqref{neumann:eq_Sdef}
\begin{equation}
	\bm{\Sigma}(z) = \textbf{A}(z)\bm{\Lambda}(z) \textbf{A}(z)^\top,
    \label{neumann:eq_spd_cond}
\end{equation}
where the matrix $\bm{\Lambda}(z) = \text{diag}(\lambda_1(z),\dots,\lambda_p(z))$ holds the conditional eigenvalues
in decreasing order and $\textbf{A}(z)=[\mathbf{a}_1(z) \ \dots \ \mathbf{a}_p(z)]$ the corresponding eigenvectors, the \emph{principal components}. The first principal components describe those processes in the data that lead to the most significant fluctuations. The eigenvalues in $\bm{\Lambda}(z)$ are the variances of the principal components. 
Once the eigenvalues and eigenvectors are estimated (denoted by $\hat{\lambda}_j$ and $\hat{\mathbf{a}}_j$), they can be used for \emph{conditional} feature extraction or data normalization. For that purpose, we extract the corresponding scores
\begin{equation}\label{neumann:eq_score_adjust}
	\textbf{s}_i = (\textbf{x}_i-\hat{\mathbf{m}}(z_i))^{\top} \hat{\textbf{A}}(z_i) (\hat{\bm{\Lambda}}(z_i))^{-1/2},
\end{equation}
with $(\hat{\bm{\Lambda}}(z_i))^{-1/2} = \text{diag}(\hat\lambda_1^{-1/2}(z_i),\dots,\hat\lambda^{-1/2}_p(z_i))$, 
$\hat{\textbf{A}}(z_i) = [\hat{\mathbf{a}}_1(z_i) \ \dots \ \hat{\mathbf{a}}_p(z_i)]$, 
and $\hat{\mathbf{m}}(z_i)$ being an estimate of the conditional mean of $\mathbf{x}$ at $z_i$ as described in Section~\ref{neumann:sec_nonoparm_est_cond_cov}. Using the conditional mean and eigenvectors and eigenvalues removes the effect of the confounder $z$ from the component scores, as those are uncorrelated, standardized quantities, each with zero mean and variance one for any given $z$-value.
Note that the conditional covariance requires the confounder to be measured; hence, conditional PCA is not an unsupervised method anymore.

\section{Proof of Concept Study}
\label{neumann:sec_validation_of_ana_method}
For proof of concept, the proposed method from Section~\ref{neumann:sec_nonoparm_est_cond_cov} is applied to artificially generated data. The purpose is to demonstrate that the confounder-adjusted covariance is estimated correctly, based on Monte Carlo simulation, and to increase the reproducibility of the results by providing a simple data set where the ground truth is known. 

\begin{figure}[!htb]
    \centering
    \includegraphics[width = .7\textwidth]{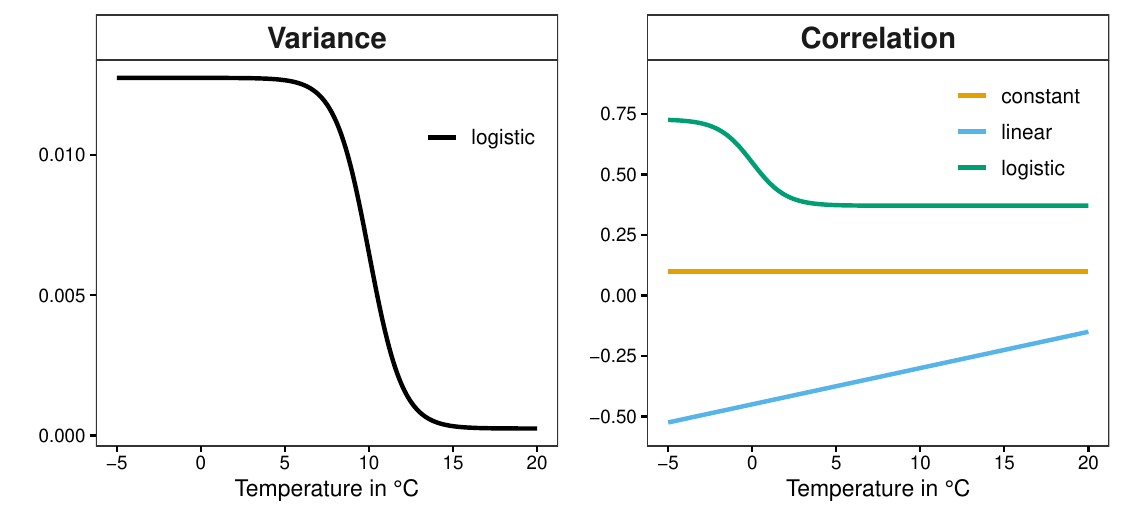}
    \caption{The (conditional) variance $\sigma^2_1(z) = \sigma^2_2(z) = \sigma^2_3(z)$ (left) and the (conditional) correlations $\rho_{1,2}(z)$, $\rho_{1,3}(z)$, and $\rho_{2,3}(z)$ (right) for the three system outputs.}
    \label{neumann:fig1}
\end{figure}

\subsection{Experimental Setup}

In this study, we considered a system with (conditionally normal) output quantities, representing measurements obtained with three different sensors. All outputs exhibit zero mean and temperature-dependent (co-)variances, as shown in Figure~\ref{neumann:fig1}. The left plot describes the main diagonals $\sigma^2_j(z)$ of the covariance matrix, $j=1,2,3$. Each of the three variance functions $\sigma^2_j(z)$ is modeled as a logistic function, where lower temperatures lead to larger variances. The right plot, on the other hand, visualizes the off-diagonal terms $\rho_{j,k}(z) = \sigma_{j,k}(z)/(\sigma_j(z)\sigma_k(z))$, $k \in \{2,3\}$, $j < k$, normalized to values in the range $[-1,+1]$. That means, correlation functions  
are shown instead of covariance functions because covariance functions can have values in the range $(-\infty, +\infty)$ and are more challenging to interpret.
The three correlation functions are modeled as follows:

\begin{itemize}[noitemsep]
    \item[(a)] \textit{Constant}. This case corresponds to the assumption that correlations between system outputs do not change with temperature, an assumption that was made in most previous publications.
    \item[(b)] \textit{Linear}. In this case, the correlation linearly decreases (in absolute value) for increasing temperatures.
    \item[(c)] \textit{Logistic}. The correlations are high for cold temperatures below $0^\circ$C, but substantially drop around $0^\circ$C and remain constant afterward. That means that the correlation is substantially larger for cooler temperatures (below $5^\circ$C) than for warmer ones.
\end{itemize}

To validate the covariance estimation method from Eq.~\eqref{neumann:eq_Sdef}, the correlation functions (a) -- (c) and the variance function should be retrieved based on the outputs and corresponding temperature readings.
To demonstrate this, 50 Monte Carlo runs were performed with varying sample sizes $n$ and different bandwidths $h$. Instead of using the conditional mean approach, a constant mean value of zero was set, as the focus in this paper is not to estimate the mean but the conditional covariance. Temperatures were assumed to be observed on an equidistant grid $\mathbf{z}_\text{seq}$, ranging from $-5^\circ$C to $20^\circ$C in steps of $0.1^\circ$C, and $n$ data points were generated per temperature value. Subzero temperatures are assumed to occur less frequently, so the samples were thinned out in the range from $-5^\circ$C to $5^\circ$C. To do so, each temperature value was assigned the probability $\pi$, where $\pi$ is $0.1$ for $-5^\circ$C and linearly increases to $1.0$ for temperatures equal to or greater than $5^\circ$C.

\subsection{Results}
\begin{figure}[!htb]
    \centering
    \includegraphics[width=.98\textwidth]{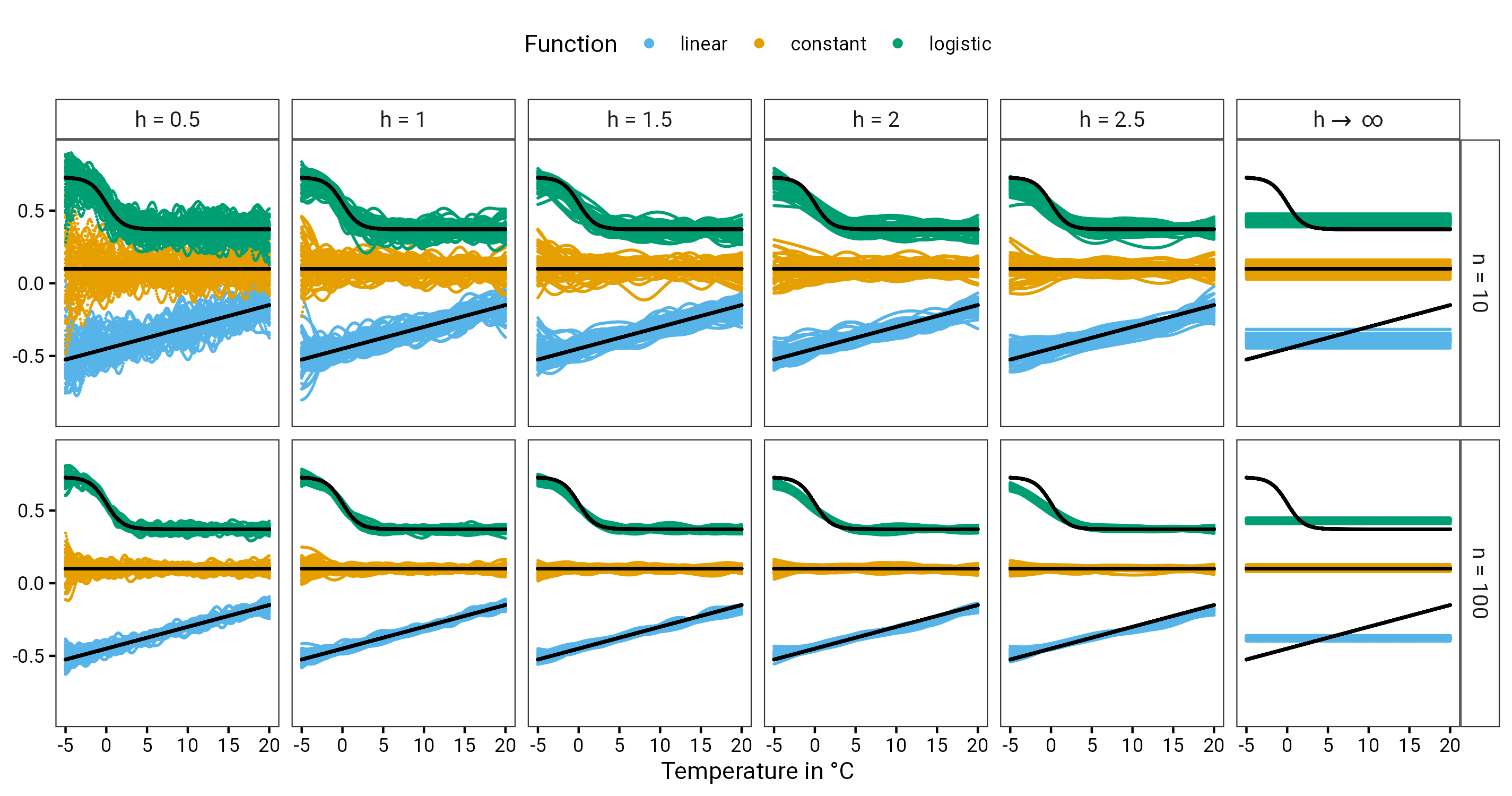}
    \caption{The estimated conditional correlations for three scenarios/sensor pairs, varying confounder values (temperature), two sample sizes (rows), and different values of the bandwidth (columns) over 50 runs of the Monte-Carlo simulation with the true correlation functions in black.}
    \label{neumann:fig2}
\end{figure}

\begin{figure}[!htb]
    \centering
    \includegraphics[width=.98\textwidth]{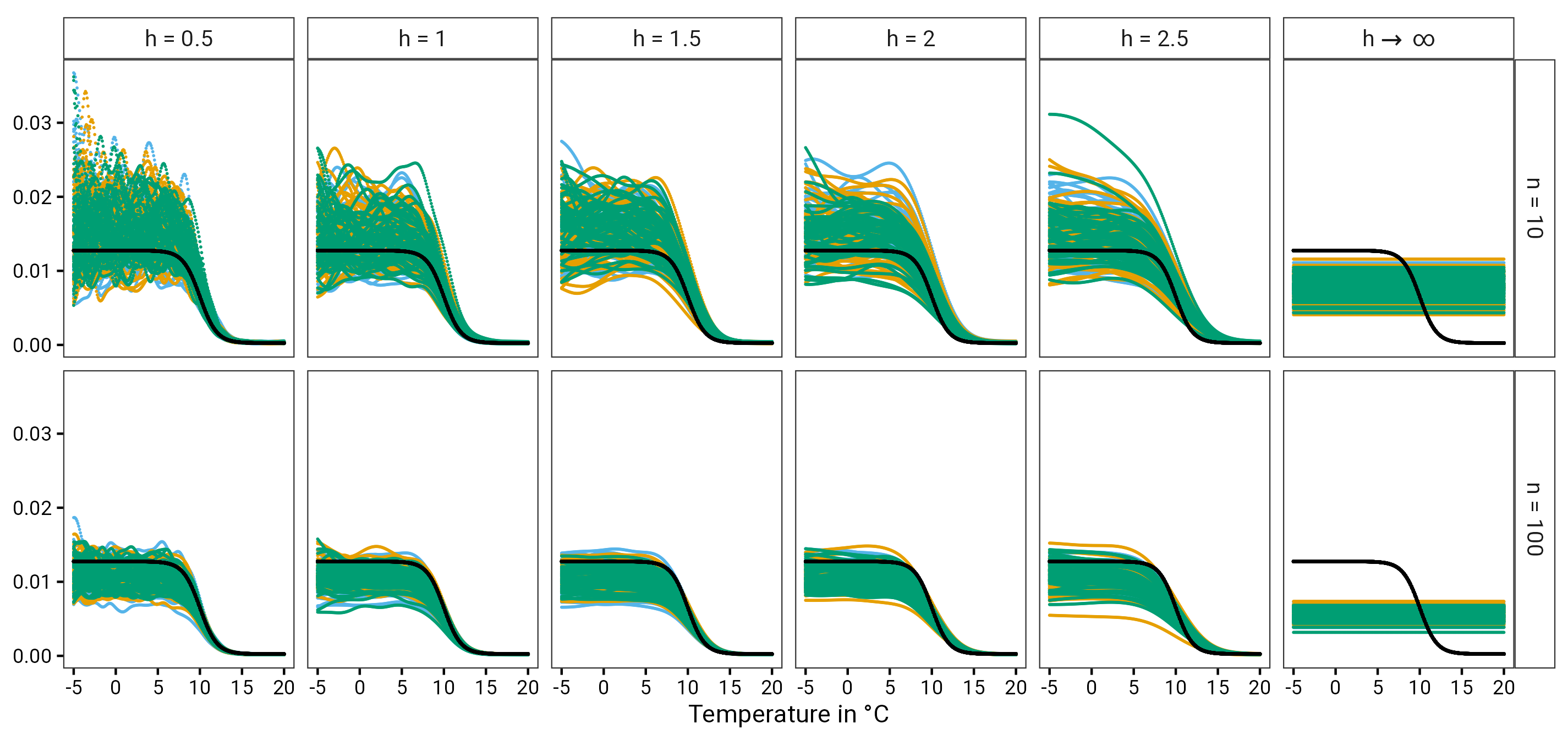}
    \caption{The estimated conditional variances for the three output quantities (blue, orange, green), varying confounder values, two sample sizes (rows), and different values of the bandwidth (columns) over 50 runs of the Monte-Carlo simulation with the true variance function in black.}
    \label{neumann:fig3}
\end{figure}

The resulting estimates of the conditional correlations are shown in Figure~\ref{neumann:fig2}, and the estimated conditional variances in Figure~\ref{neumann:fig3}. The graphs are divided according to the user-defined bandwidth $h$ (columns) and the sample sizes (rows). The figures show the estimated conditional correlations and conditional variances, respectively, for all four target functions.
The result of each Monte Carlo run is plotted as a separate function, and the target functions from Figure~\ref{neumann:fig1} are shown as well for comparison (solid black lines).
The last column in Figure~\ref{neumann:fig2} and \ref{neumann:fig3} shows the results for the standard approach assuming a constant covariance, that is not affected by the temperature, which corresponds to the extreme case of bandwidth $h\rightarrow\infty$.


We can appreciate that the non-parametric approach proposed in Section~\ref{neumann:sec_nonoparm_est_cond_cov} approximates the true functions well, but the accuracy of the estimation depends on the chosen bandwidth. On one hand, the estimates of the target function become smoother for higher bandwidths and larger sample sizes. On the other hand, a chosen bandwidth that is too large results in substantial bias if the true correlation indeed depends on the temperature (`over-smoothing'). In other words, the larger the bandwidth, the more the estimated correlation function is pushed toward a constant value, in particular for the extreme case with $h \rightarrow \infty$. Consequently, the estimate is biased if the true correlation is not constant. We can also see that the estimates exhibit a larger variance for smaller bandwidths (`under-smoothing'), which is particularly visible for the smaller sample size with $n=10$, and ultimately, we can observe an even more pronounced scatter for negative temperatures due to the thinned-out data sets. The two main takeaways are that the proposed non-parametric approach can identify confounder effects on conditional covariances and that it is important to choose an appropriate value for the smoothing parameter. 

\begin{figure}[!htb]
    \centering
    \includegraphics[width=\textwidth]{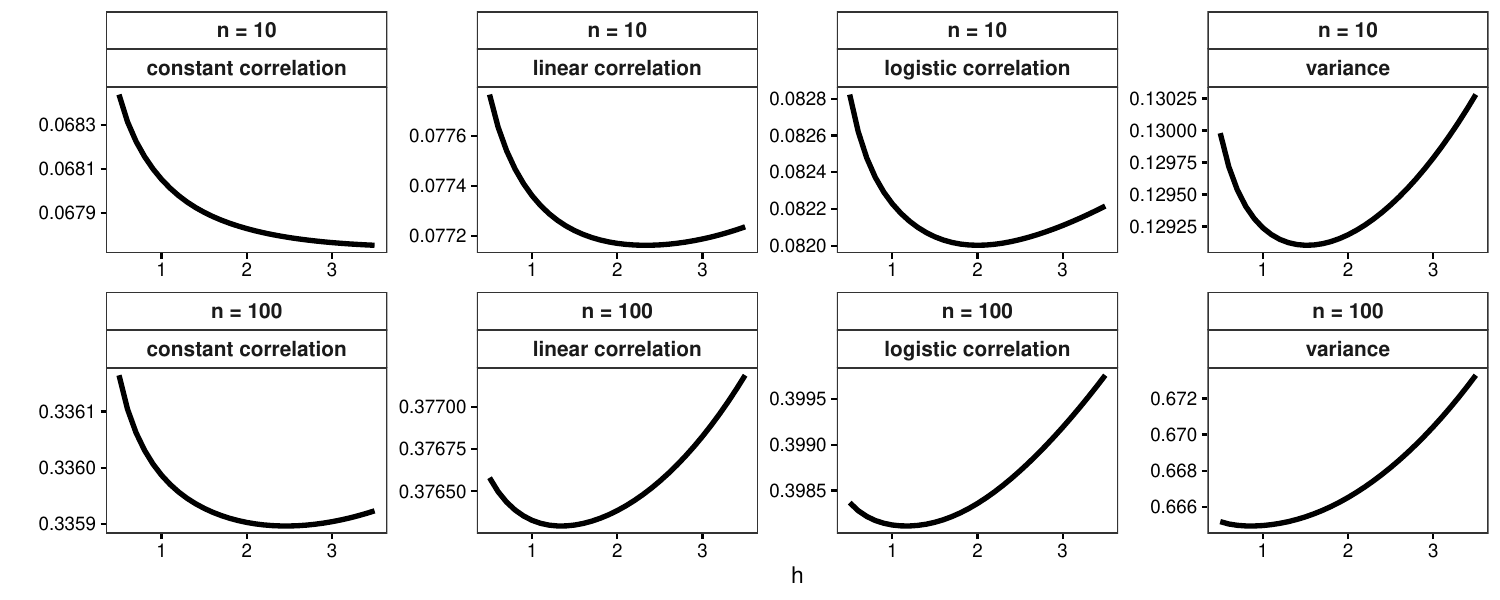}
    \caption{Optimizing the bandwidth parameter based on the quadratic loss from Eq.~\eqref{neumann:eq_QSig} for different target functions and sample sizes.}
    \label{neumann:fig4}
\end{figure}

To find the optimal bandwidth, the loss function $Q_{\sigma_{j,k}}$ from Eq.~\eqref{neumann:eq_QSig} is suggested. The optimal value is chosen as the minimum in the loss function, and Figure~\ref{neumann:fig4} displays the loss function for different bandwidths, sample sizes (rows), and target functions (columns). On close inspection, it can be seen that, for a constant correlation function, the optimum is achieved for a very large bandwidth. This is because, with the bandwidth tending towards infinity, the estimated correlation becomes a constant and thus independent of the confounder. 
For the linear case, a large bandwidth of about 2.5 produces good results for a small sample size, but a smaller bandwidth of about 1.5 would have been more appropriate for a larger sample size, see Figure~\ref{neumann:fig4}. These values indeed led to the optimal fits in Figure~\ref{neumann:fig2} (blue curves), which proves the effectiveness of the suggested optimization procedure. 
For the logistic correlation function, the curves in Figure~\ref{neumann:fig4} have well-defined minima as well. With $n=10$, the optimal bandwidth is about $h=2$, but it reduces to $h\in [1,1.5]$ for a larger sample size of $n=100$. Comparing these results with the green curves in Figure~\ref{neumann:fig2} confirms the quality of the fit. 
Ultimately, the results for the variance function are similar with distinct minima at $h=1.5$ and $h=1$ depending on the sample size.

This concludes the validation study based on artificially generated data. The results showed that the proposed method could retrieve the underlying target functions well and that the suggested procedure for selecting the bandwidth parameter(s) leads to plausible results. This increases the credibility of the developed approach and allows us to move on to more complex case studies with real-world data.

\section{Application to Real-World SHM Data}\label{neumann:sec_application_SHM_data}
In this section, the proposed method from Section~\ref{neumann:sec_method} and \ref{neumann:sec_shm}
will be applied to data sets from two real-world bridges. Measurements from different types of sensors (strain gauges, inclinometers, and accelerometers) and extracted natural frequencies are analyzed to demonstrate the versatility of the developed methods. The examined case study includes a composite bridge and a steel railway bridge with monitoring durations between 22 days and 7.5 months, respectively. 

\subsection{Munich Test Bridge}
\label{neumann:sec_tbm}
\begin{figure}[h]
\centering
    \includegraphics[width = .38\textwidth]{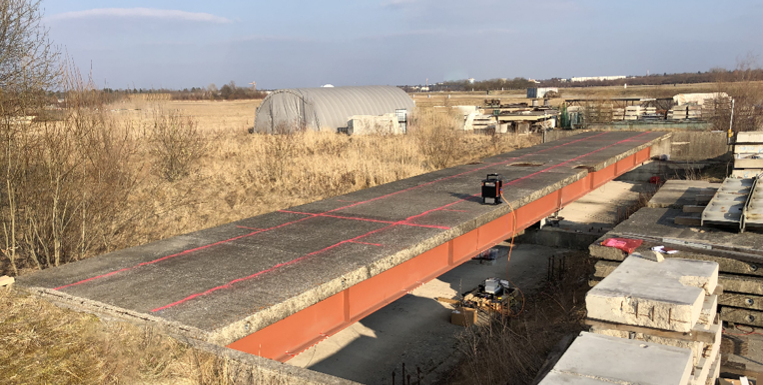} 
    \includegraphics[width = .6\textwidth]{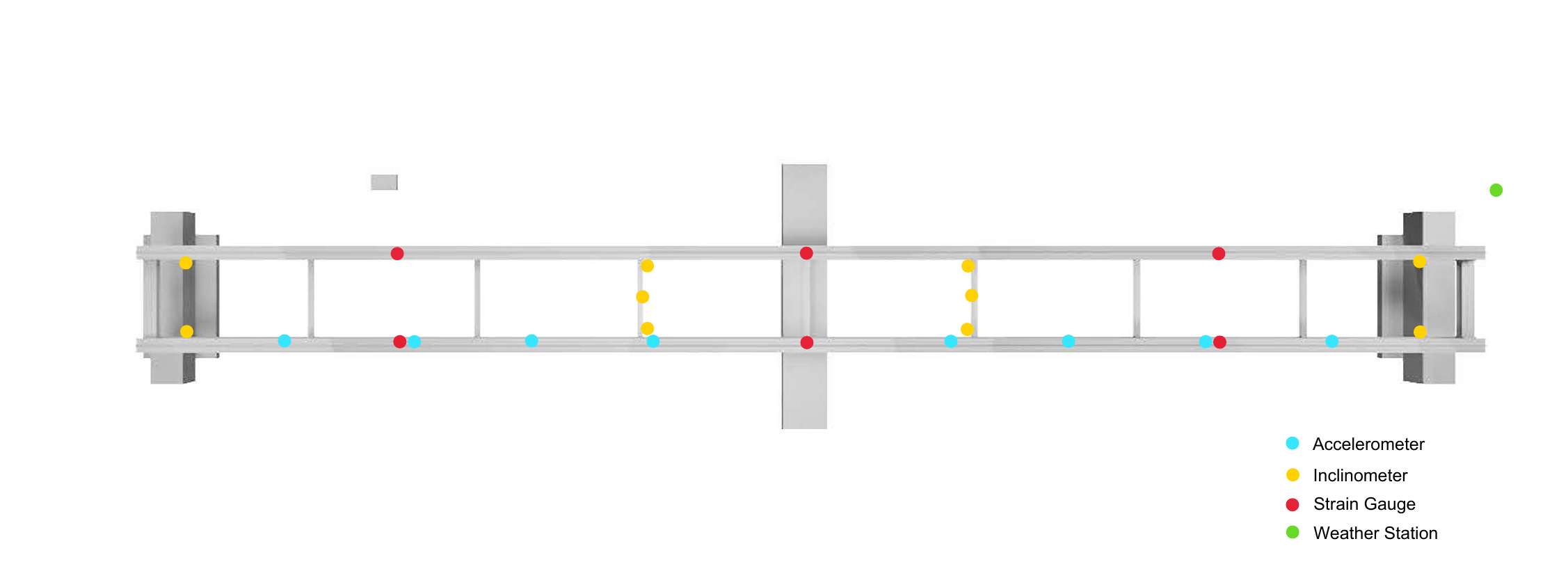}
    \caption{Photo (left) and instrumentation plan (right) of the Munich Test Bridge.}
    \label{neumann:fig5}
\end{figure}

The first case study is the Munich Test Bridge, a 30-meter-long steel-concrete composite bridge on the University of the Bundeswehr Munich grounds. It was built in 2007, and a photo and a schematic drawing can be seen in Figure~\ref{neumann:fig5}. 
The bridge deck consists of 11 concrete segments with varying lengths, a width of 4 meters, and a slab thickness of 0.2 meters. The underlying steel component consists of two longitudinal steel girders, seven transverse braces (at Axes 2 to 8), and two transverse braces at the abutments. A joint group from Helmut Schmidt University, the University of the Bundeswehr Munich, and the Technical University of Munich (TUM) collected measurement data between March 4, 2022, and April 11, 2022. The data set consists of three weeks of reference measurements and four damage scenarios (extra masses, bolt failure, brace failure, support settlements), described in detail by \citet{Jaelani.etal_2023}. In this paper, we focus on the reference measurements, as they demonstrate the effect of varying ambient temperatures on the output covariances while the bridge remains undamaged. In particular, acceleration measurements from eight vertical accelerometers are processed, as well as the measurements from ten inclinometers and strain measurements from six strain gauges, see Figure~\ref{neumann:fig5}. Table~\ref{neumann:tab_samplingfreq_munich} shows all other signal processing parameters. When analyzing acceleration measurements, temperature measurements are interpolated to ensure the same resampling frequency of 100~Hz, using the \texttt{signal} R package~\citep{Rsignal_2014}. 

\begin{table}[!h]
    \centering
    \tabcolsep7.1pt 
    \small
    \caption{Signal processing parameters for the Munich Test Bridge.}
    \begin{tabular}{l *{4}{c}}
	\toprule
	\multicolumn{1}{c}{\textbf{Sensor Type}} & 
        \multicolumn{1}{c}{\textbf{Sampling}} & 
        \multicolumn{1}{c}{\textbf{Resampling}}& 
        \multicolumn{1}{c}{\textbf{Selected}} &
        \multicolumn{1}{c}{\textbf{Resulting}}\\
	\multicolumn{1}{c}{} & 
        \multicolumn{1}{c}{\textbf{Frequency (Hz)}} & 
        \multicolumn{1}{c}{\textbf{Frequency (Hz)}}& 
        \multicolumn{1}{c}{\textbf{Bandwidth}}&
        \multicolumn{1}{c}{\textbf{Sample Size}}\\
	\midrule
	Acceleration & 1000 & 100 & $h = 0.5$ & $n=178$,$740$,$000$\\
	Inclination & 100 & 1 & $h \in [0.5,2.5]$ & $n=1$,$785$,$480$ \\
	Strain & 100 & 1 & $h \in [0.8,2.5]$ & $n=1$,$785$,$480$\\
	Temperature & 1 & 1 & & $n=1$,$785$,$480$\\
	\bottomrule
    \end{tabular}     \label{neumann:tab_samplingfreq_munich}
\end{table}
%

\subsubsection{Temperature Effects on Covariances} \label{neumann:sec_cond_covTBM}
The following results clearly show that the output covariances depend on temperature readings.  
The non-parametric, kernel-based estimator is employed to estimate the conditional mean and the conditional (concurrent) covariances according to Eq.~\eqref{neumann:eq_mdef} and \eqref{neumann:eq_Sdef}, respectively. The bandwidth $h_{j,k}$ for each sensor pair $j,k$ was determined by dividing the data into a training and validation set and by minimizing the squared loss $Q_{\sigma_{j,k}}(h)$, Eq.~\eqref{neumann:eq_QSig}, separately for each sensor pair and sensor type as described in Section~\ref{neumann:sec_nonoparm_est_cond_cov}. The optimal bandwidth varies depending on whether acceleration, strain, or inclination data is examined; see Table~\ref{neumann:tab_samplingfreq_munich}. For acceleration channels, the chosen bandwidths are comparatively small (presumably due to the large sample size/resampling frequency, compare Section~\ref{neumann:sec_validation_of_ana_method}), while those for strain channels are higher. For inclination channels, the optimal bandwidths are distributed almost uniformly between 0.5 and 2.5. Similar to the trends in Figure~\ref{neumann:fig2} and Figure~\ref{neumann:fig3}, however, the results appear relatively robust for different bandwidth choices.
Fortunately, the temperature differences within the three-week measurement period and within a day (day and night difference) were very large, so we could estimate the conditional covariances for temperature values between $-5^\circ$C and $20^\circ$C.%

\begin{figure}[!htb]
    \centering
    \includegraphics[width = \textwidth]{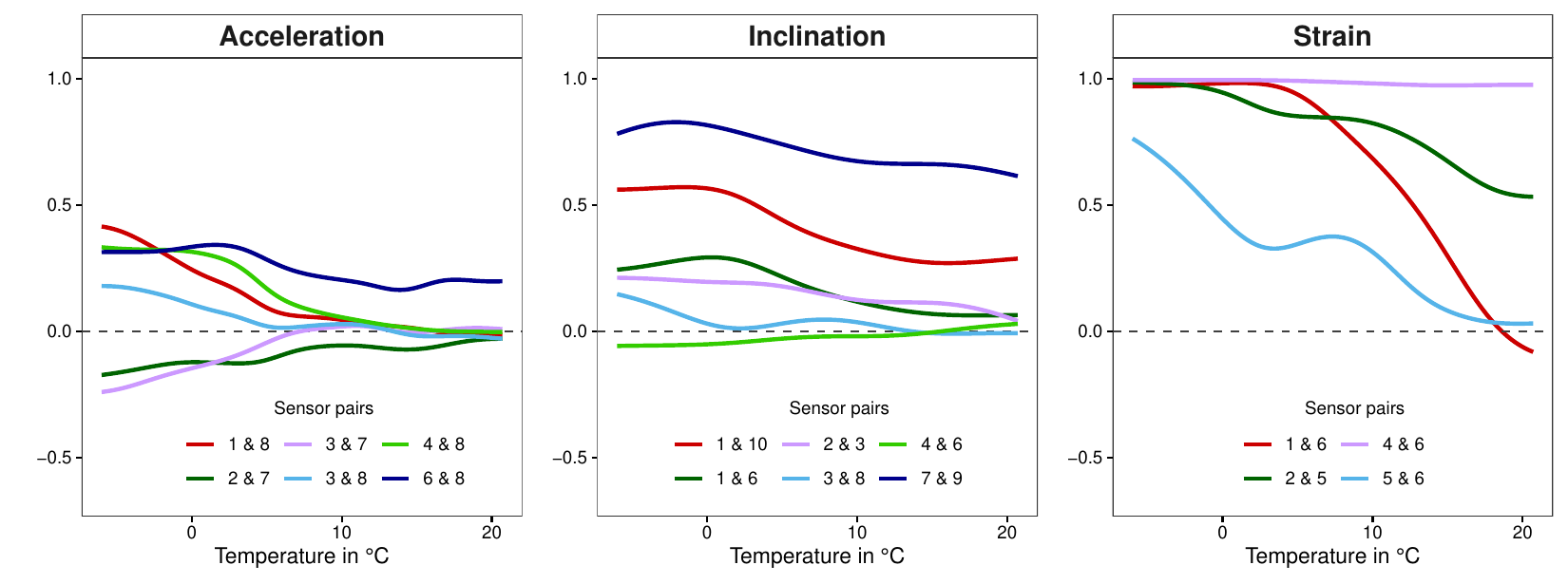} 
    \caption{Non-parametric estimates of the conditional (concurrent) correlation of sensor pairs for the test bridge's acceleration, inclination, and strain measurements as a function of temperature $z$.}
    \label{neumann:fig6}
\end{figure}

The results from the analysis are covariance matrices with entries that depend on temperature. Those matrices are challenging to visualize for varying temperatures. That is why only selected matrix entries are plotted as functions of temperature for acceleration, inclination, and strain measurements in Figure~\ref{neumann:fig6}. 
Again, correlation functions are shown instead of covariance functions because the former are more straightforward to interpret due to their normalization to the $[-1,+1]$ range (cf. Section~\ref{neumann:sec_validation_of_ana_method}). The most important finding is that almost all correlation functions are affected by temperature changes, which justifies the methodological contributions in this paper. The impact is especially noticeable for low temperatures, where most of the correlation functions from Figure~\ref{neumann:fig6} reach their maximum, and the correlation tends to reduce as temperature increases. The changes in the correlation functions are more pronounced for acceleration and strain measurements than for inclination measurements, meaning that, in the case of the Munich Test Bridge, conditional covariances are particularly important for acceleration and strain sensors.

\subsubsection{PCA-based Feature Extraction} \label{neumann:sec_conf_adj_PCA_via_cond_covTBM}
\begin{figure}[!htb]
	\centering
	\includegraphics[width=\textwidth]{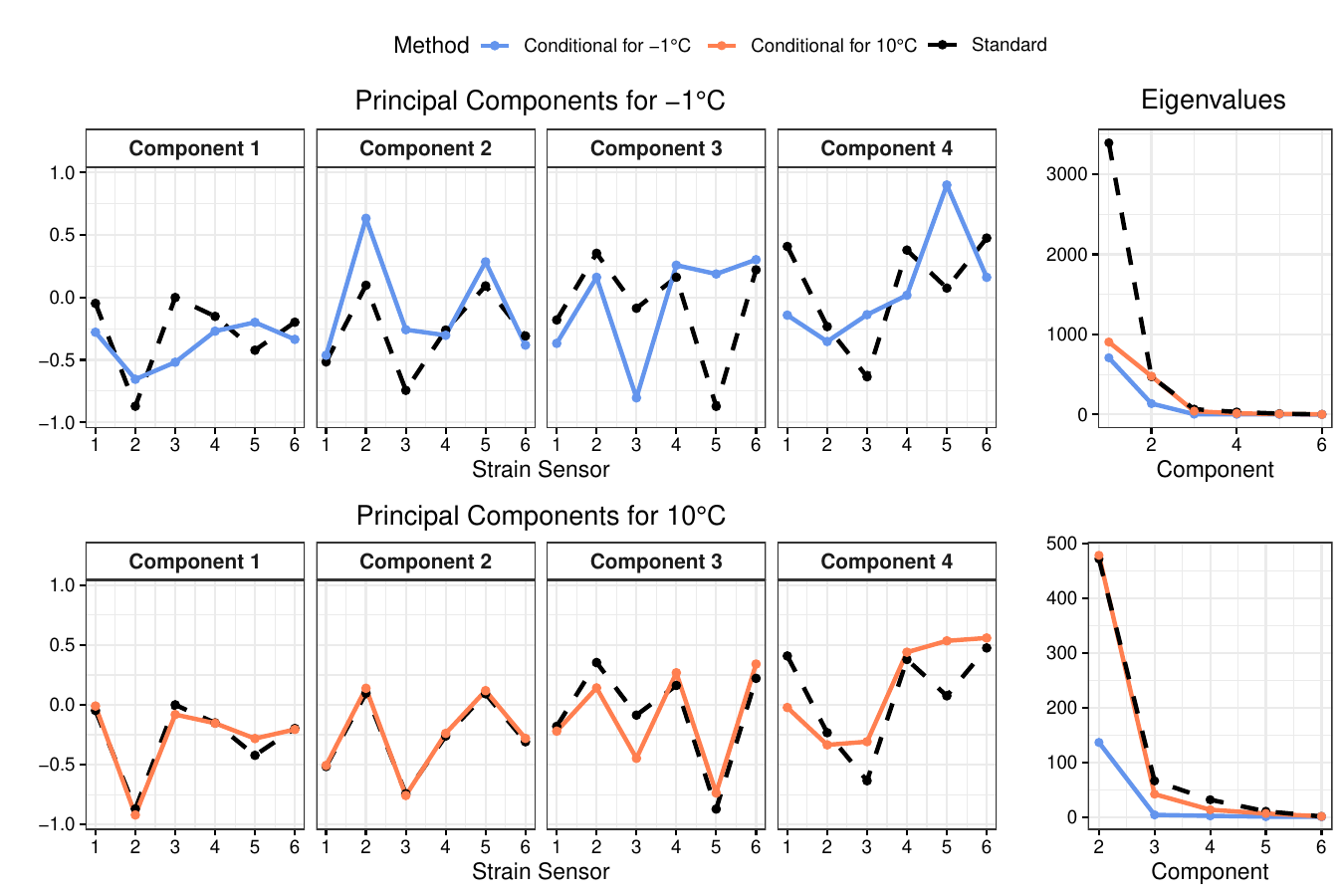} 
	\caption{Standard principal components (black) and conditional principal components for the strain data of the Munich test bridge and two different temperature values with $-$1°C (blue) and 10°C (orange), together with the corresponding eigenvalues.}
	\label{neumann:fig7}
\end{figure}

In this section, PCA is applied to highlight that the developed approach can be used to extract features that are robust toward temperature changes. In Figure~\ref{neumann:fig7}, the first four conditional principal components for strain measurements are given for temperature $z=-1^\circ$C (blue) and $z = 10^\circ$C (orange), and the standard principal components using the marginal covariance (dashed black). For conciseness, we limit the illustration to the first four principal components, as they explain 99.7\% of the total variance (according to the standard approach). The conditional PCA shows that both the direction and the variance of the principal components change with changing temperature. If the standard version were used, this fact would be disregarded.
Particularly for the subzero temperature value of $-$1°C, the conditional principal components are distinctly different from the standard ones (as seen from the top left subplots of Figure~\ref{neumann:fig7}). In other words, for low temperatures, the main directions of the variation in the strain measurements are substantially different from those we would (wrongly) obtain from standard PCA, ignoring potential temperature effects on covariances. 
In the right column of Figure~\ref{neumann:fig7}, the eigenvalues are shown, with the bottom plot showing the second to 6th eigenvalues only. The effect of changing (co-)variances becomes very obvious in the eigenvalues, as they differ substantially between cold (blue) and warm (orange) temperatures. The finding that, in particular, the first eigenvalue is much larger for standard PCA than for conditional PCA can be explained by the fact that conditional PCA removes the temperature-induced variation from the data by using the conditional mean instead of the marginal one. In summary, the principal components change substantially due to changes in the covariances, particularly at lower temperatures. If this effect is not accounted for, it will not be possible to remove the impact of confounding variables from the data or to distinguish between temperature changes and structural damage. 
If confounder information is available (e.g., temperature readings), PCA can be modified to exploit this additional information, and conditional component scores that are robust to changes in the confounding variables can be calculated, as will be shown next. 

\begin{figure}[!htb]
    \centering
    \includegraphics[width = .95\textwidth]{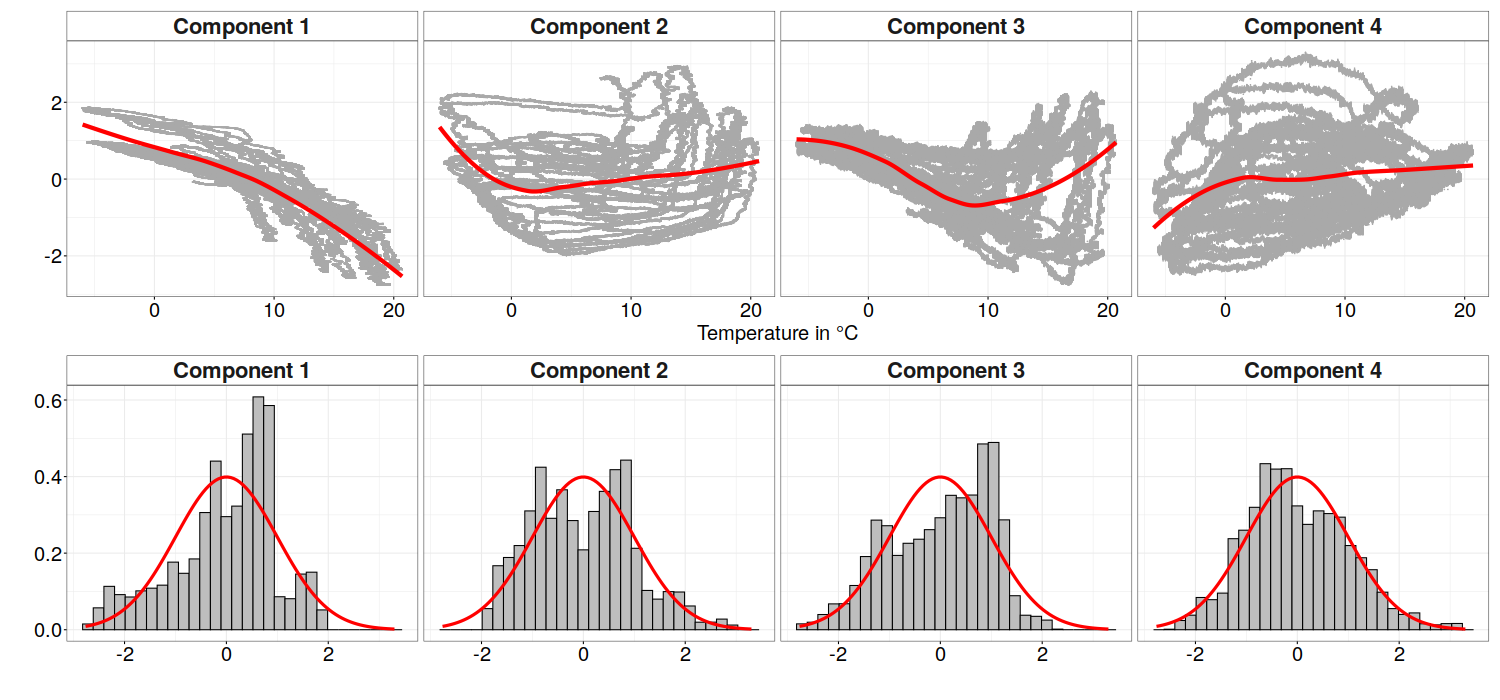}
    \caption{Standardized scores from conventional PCA (grey dots, top panel) and their (conditional) mean (red, top panel) as a function of temperature, together with the distribution of the scores (grey histograms, bottom panel) in comparison to the standard normal density (red, bottom panel).}
    \label{neumann:fig8}
\end{figure}

\subsubsection{PCA-based Temperature Elimination}
PCA is a suitable approach to remove temperature effects from arbitrary system outputs, see Section~\ref{neumann:sec_pca}. To do so, the user can reject some principal components (sources of variance), and reconstruct the system outputs based on the remaining ones that are not affected by confounders. Existing publications suggest that removing one or two principal components suffices to remove one confounder, but this section will challenge this presumption.

\begin{figure}[h]
    \centering
    \includegraphics[width = \textwidth]
    {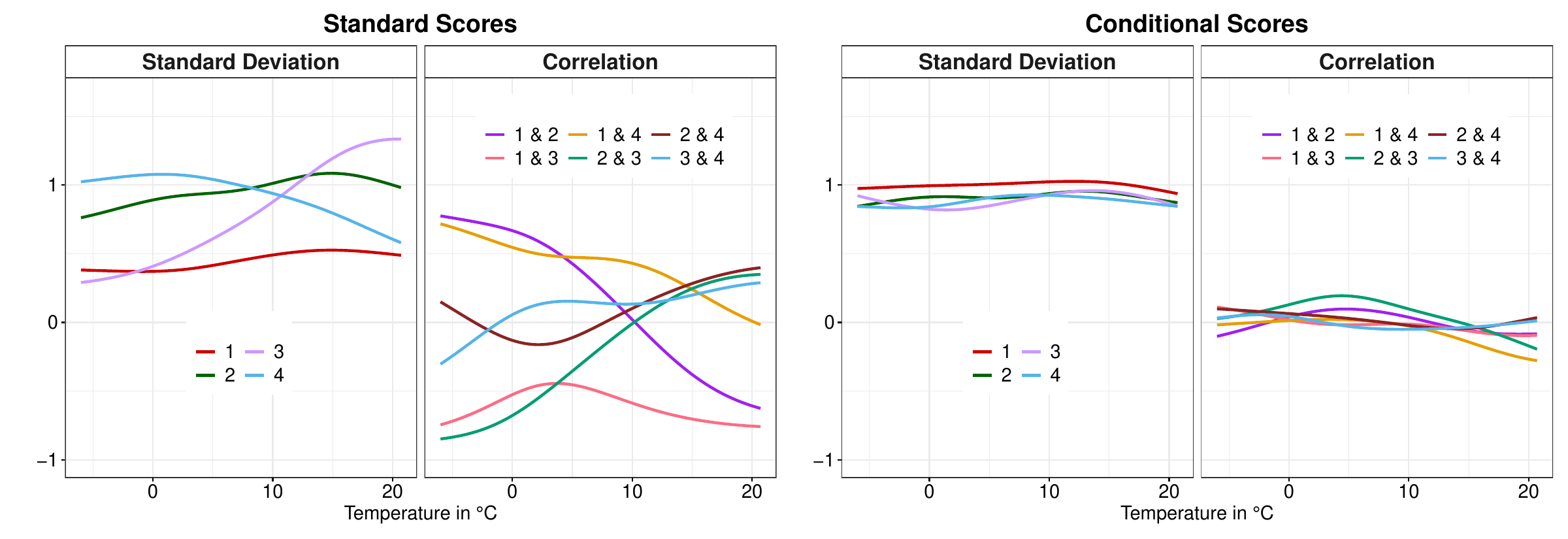}
    \caption{Standard deviations and correlations of the standard PC scores (left) and conditional scores (right) for the strain data from the Munich test bridge.}
    \label{neumann:fig9}
\end{figure}

For demonstration, the scores from Eq.~\eqref{neumann:eq_score_adjust} will be plotted as a function of temperature for conventional PCA and the developed PCA approach. The results for conventional PCA are shown in Figure~\ref{neumann:fig8}.
The top subplot clarifies that the standard scores of the first principal component exhibit a strong and almost linear association with temperature. This was expected as it has been reported earlier that the first principal component(s) mainly describe(s) confounder-induced variation, cf.~\citet{Cross.etal_2012}. However, the second to fourth components also show a clear and non-linear relationship with temperature.
The bottom subplot of Figure~\ref{neumann:fig8} shows the marginal distributions of the standard scores (grey histograms) together with the theoretical distributions, that is, standard normal distributions (red). It can be observed that the scores deviate substantially from the normal distribution for the first four components, meaning the first four scores are affected by temperature changes.
To further study this, the standard deviations of the standard scores and their correlations are shown as a function of temperature in Figure~\ref{neumann:fig9}~(left).
Most of the correlations and some of the standard deviations vary substantially with changing temperature and do not approximate the theoretical values of zero and one across the temperature spectrum.
So, temperature changes affect not only the distribution of the first scores but also the first three to four scores. Moreover, normality cannot be assumed.

\begin{figure}[!htb]
    \centering
    \includegraphics[width = .95\textwidth]{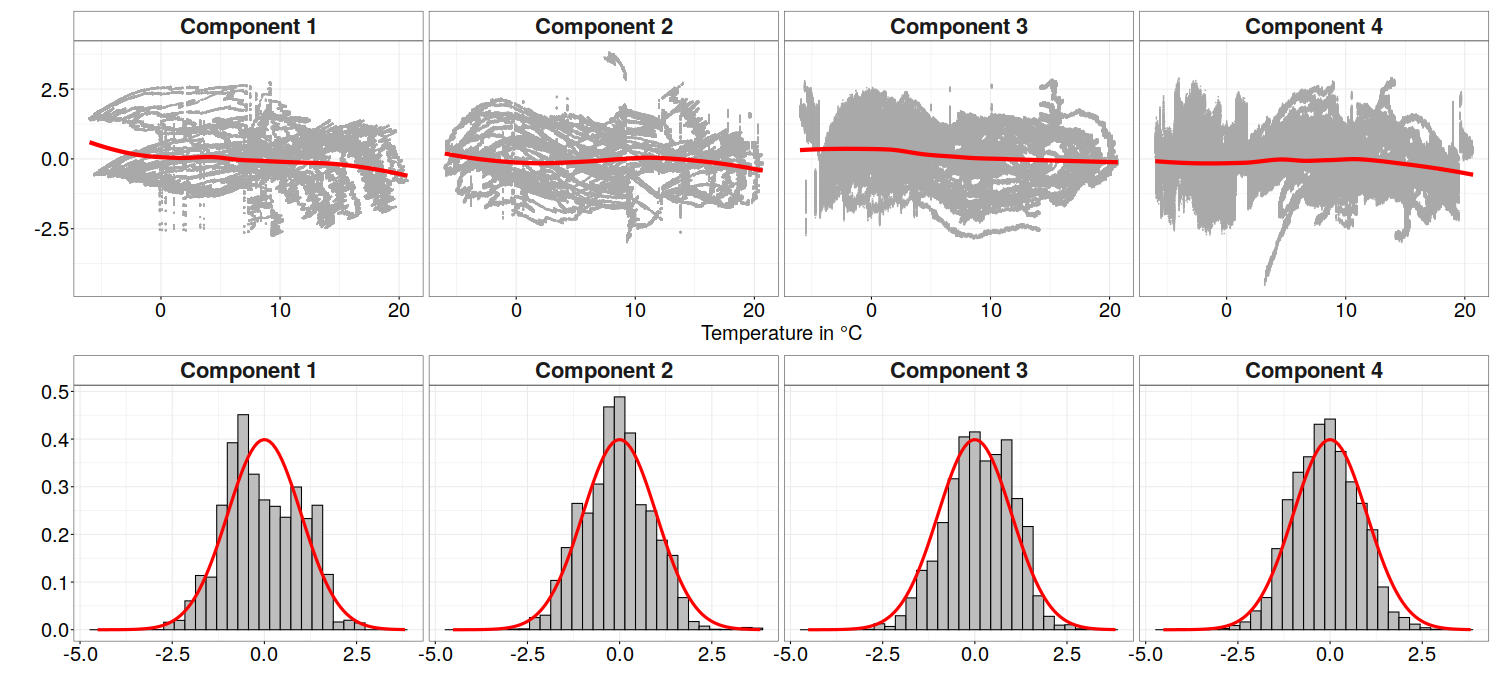}
    \caption{Standardized conditional scores (grey, top panel) and their (conditional) mean as a function of temperature (red, top panel) for the first four principal components together with histograms (grey, bottom panel) of the score values in comparison to the standard normal density (red, bottom panel).}
    \label{neumann:fig10}
\end{figure}

Repeating the analysis for the newly developed PCA, and plotting the scores as a function of temperature leads to more convincing results, as shown in Figure~\ref{neumann:fig10}. 
In contrast to the standard scores, the mean of the conditional scores is zero across the entire temperature range for all four components (top plot), indicating no association between temperature and the scores.
The bottom subplot clarifies that the scores of the second to fourth components closely follow a normal distribution. Only the first component exhibits a bimodal distribution, but this may be due to a second, unobserved confounder.
Figure~\ref{neumann:fig9}~(right) confirms that the conditional scores exhibit constant standard deviations and correlations close to zero for the entire temperature range.

In summary, for conventional PCA, the distribution of the first three to four scores is affected by temperature changes. Only with the help of the newly developed conditional PCA, it is possible to fully remove the effect of temperature changes on the mean and the (co-)variance of the scores. 
Consequently, rejecting the first conditional score and reconstructing the system outputs based on the remaining conditional scores would have been sufficient to remove the effect of temperature changes on the mean and the covariance of the outputs. What is more, if a damage detection tool relies on the assumption of normally distributed scores, conventional PCA might have led to unreliable results and an increased number of false alarms for certain temperatures.

\subsection{KW51 Railway Bridge}
\label{neumann:sec_kw51}
\begin{figure}[!htb]
    \centering
    \includegraphics[width=.64\textwidth]{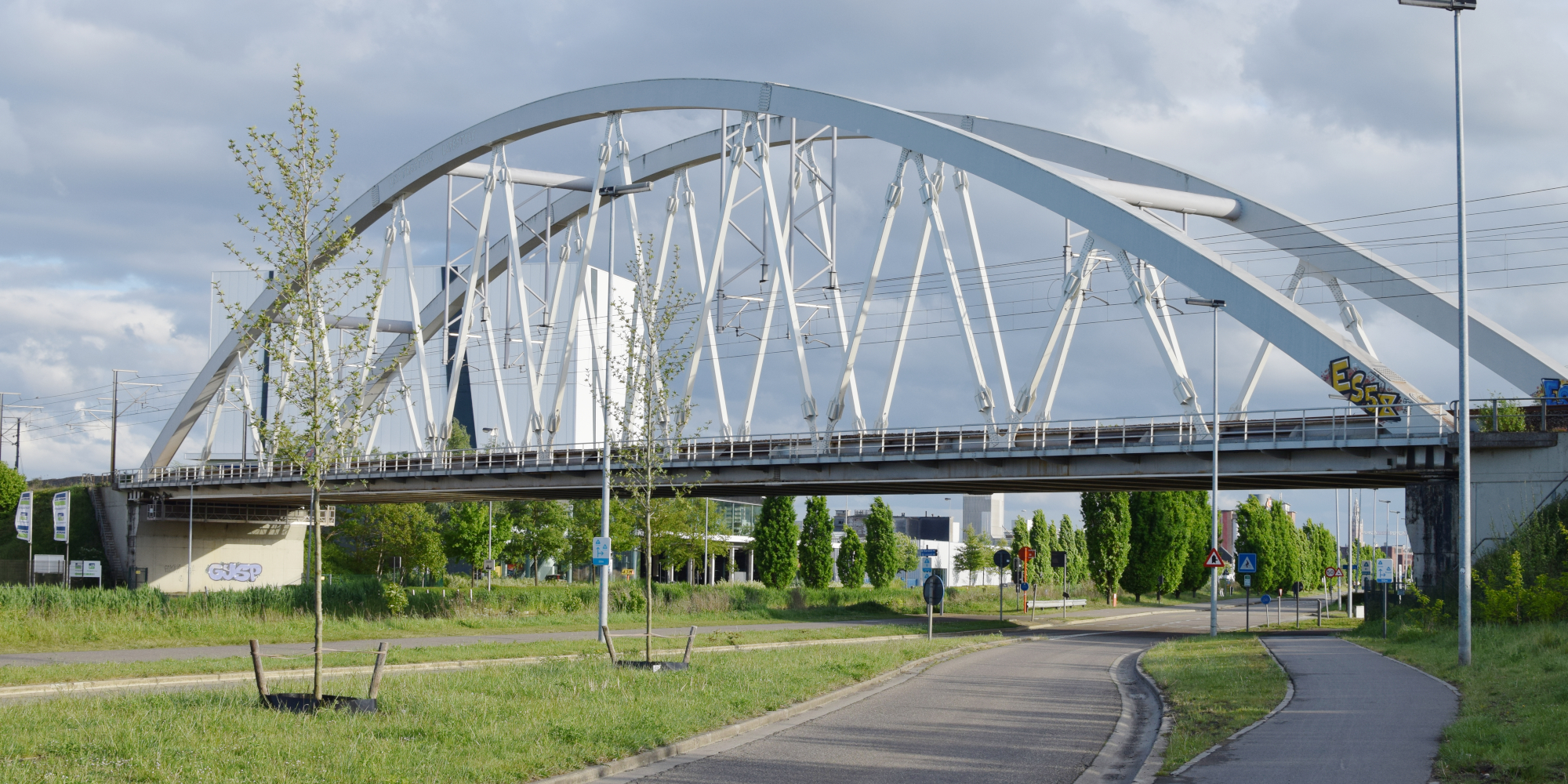}
    \includegraphics[width=.34\textwidth]{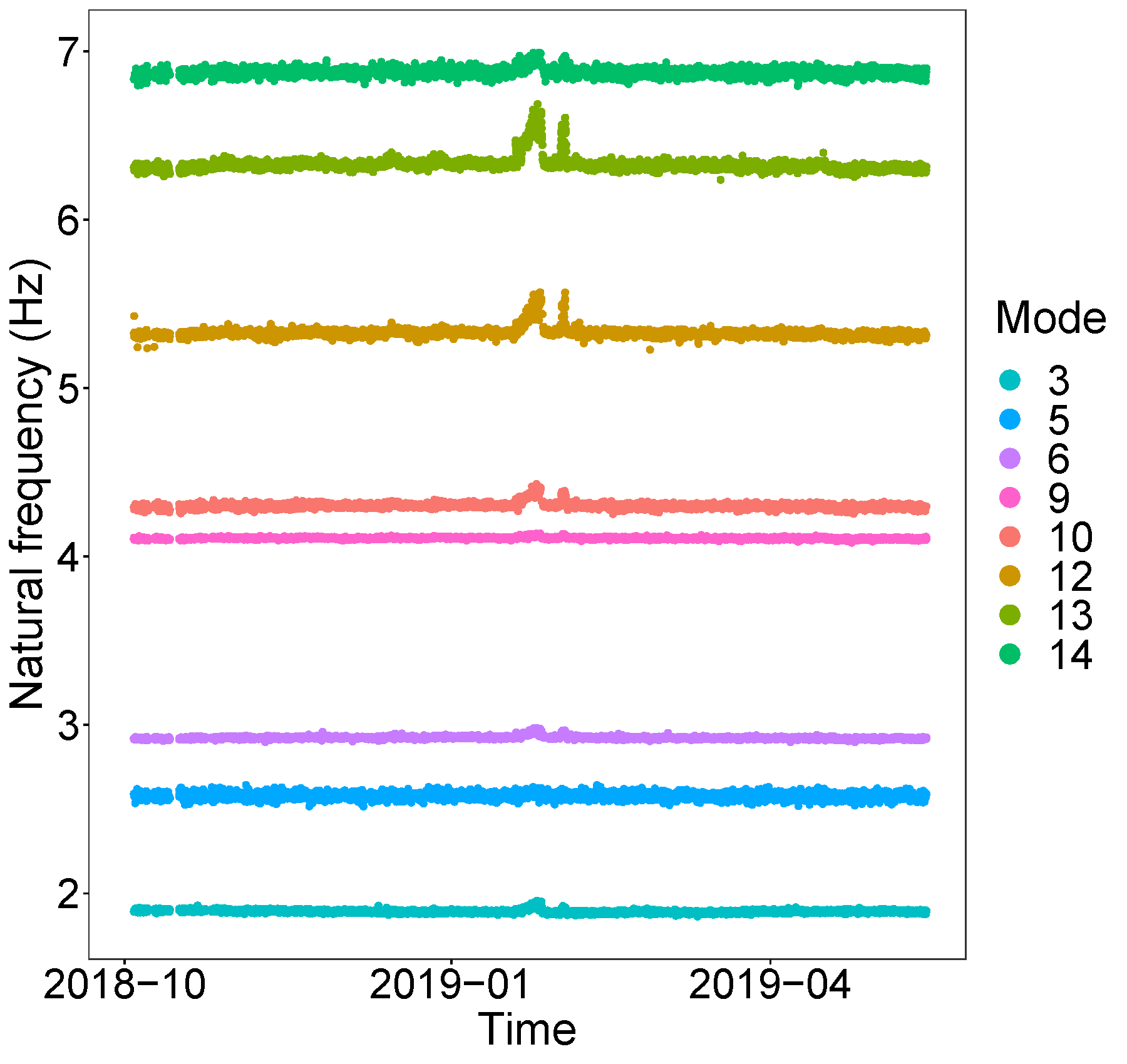}
    \caption{KW51 bridge from the north side \citep{Anastasopoulos.etal_2022} and the considered natural frequency data.}
    \label{neumann:fig11}
\end{figure}

The second case study is the KW51 Railway Bridge in Leuven \citep{Maes.Lombaert_2021}.
It is a steel railway bridge of the bowstring type; see Figure~\ref{neumann:fig11}~(left). 
It has a length of 115 meters, a width of 12.4 meters, and consists of two electrified tracks, both of which are curved. It is located between Leuven and Brussels on the railway line L36N. The railway bridge was monitored from October 2, 2018, to January 15, 2020, with a retrofitting period from May 15 to September 27, 2019. We will first have a look at the data before retrofitting, a period of 7.5 months. The sampling frequency is $1651.6$~Hz for acceleration and inclination, and the steel surface temperature is measured once per hour \citep{Maes.Lombaert_2020, Maes.Lombaert_2021, Maes.etal_2022}. 
Due to the large amount of data, only fragments of the raw acceleration and inclination data are available online~\citep{Maes.Lombaert_2020}. However, \citet{Maes.Lombaert_2021} determined modal parameters on an hourly basis
using operational modal analysis, more precisely, reference-based covariance-driven stochastic subspace identification (SSI-Cov-Ref), tracked 14 natural frequencies over time, and made them publicly available for the entire monitoring period. Hence, we will consider those natural frequencies for further analysis here and will particularly look at the conditional covariances and correlations for a given steel surface temperature. We use eight modes of vibration for our analysis (Modes 3, 5, 6, 9, 10, 12, 13, and 14) as shown in Figure~\ref{neumann:fig11}~(right). The other six modes were excluded, as the modes are often not sufficiently excited and could not be tracked reliably. For the eight considered modes, missing data was supplemented using linear interpolation as done in previous studies \citep{Maes.etal_2022}.

\subsubsection{Conditional Covariances}
In this paper, we want to focus on the estimation of the (conditional) covariance. For estimating the conditional mean, we used a bilinear model with a breakpoint of $2^\circ$C to capture the behavior of the eigenfrequencies around zero degrees analogously to \citet{Maes.etal_2022}. The conditional covariance was estimated as in Eq.~\eqref{neumann:eq_Sdef} for a temperature range between $-3^\circ$C and $26.3^\circ$C. As explained in Section~\ref{neumann:sec_bandwidth}, the bandwidth parameters were chosen by dividing the data into a training and a validation set and subsequently minimizing the squared loss from Eq.~\eqref{neumann:eq_QSig} separately for each frequency pair. The optimal bandwidth for approximately half of the mode pairs was between $0.6$ and $1.9$, and for the other half, it was around $0.5$. However, the results appeared robust against the exact choice of the smoothing parameters, similar to the trends in Figure~\ref{neumann:fig2} and \ref{neumann:fig3}.

\begin{figure}[th]
    \centering
    \includegraphics[width = .8\textwidth]{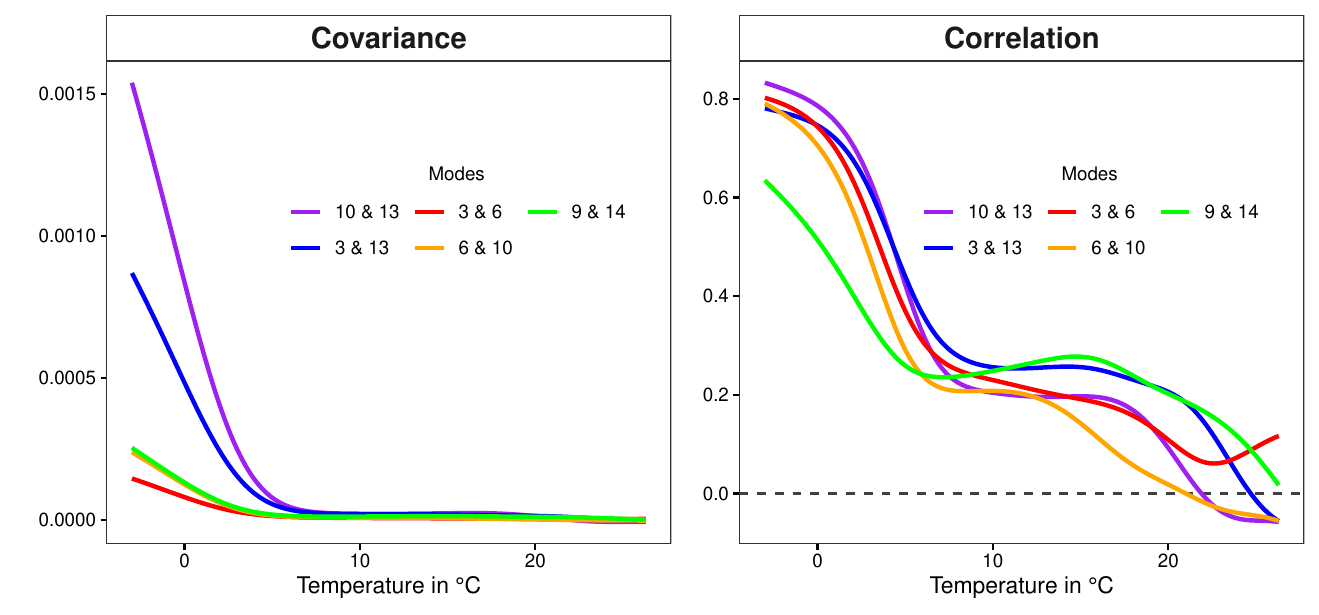}
    \caption{Non-parametric estimates of the conditional covariances (left) and correlations functions (right) for the natural frequencies of the KW51 Bridge for selected mode pairs as a function of temperature $z$.}
    \label{neumann:fig12}
\end{figure}

\begin{figure}[!htb]
    \centering
    \includegraphics[width = .9\textwidth]{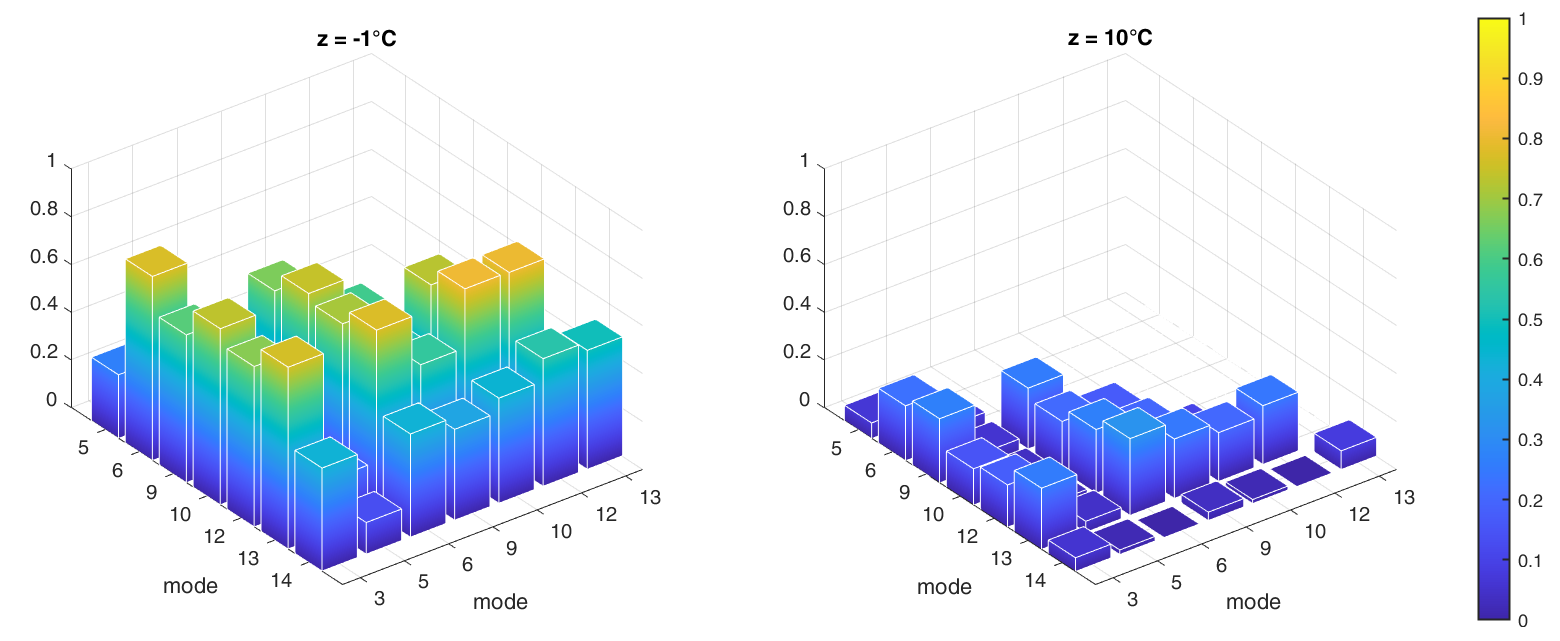}
    \caption{Visualization of the correlation matrix 
    for natural frequencies of the KW51 bridge for $-1^\circ$C (left) and $10^\circ$C (right). The matrix is symmetrical, and only the lower triangular part is shown.}
    \label{neumann:fig13}
\end{figure}

Figure~\ref{neumann:fig12} shows the conditional covariances (left) and conditional correlations (right) for five different mode pairs, so again, only some entries of the covariance or correlation matrix are shown. The maximum magnitude of the conditional covariance is $10^{-3}$ at $z = -3^\circ$C, which is very small, and for increasing temperatures, it decreases further and approaches zero. 
This effect can also be seen in Figure~\ref{neumann:fig13}, which visualizes the correlation matrix in a 3-D bar plot. 

On the left side of the figure, the lower triangular of the conditional correlation matrix is shown at $z=-1^\circ$C, and $z=10^\circ$C on the right side. This clarifies that not only the expected values of the eigenfrequencies are affected by temperature but also (co-)variances and correlations. The latter is impossible to account for by standard response surface modeling, such as the often-used (bi-)linear approach or any other method for mean regression. This underscores the relevance of the methodological developments in this paper. Obviously, the conditional covariance approach is more accurate in capturing the uncertainties in the system outputs at varying temperature values, which inevitably affects the performance of any covariance-based damage detection algorithm, but more about this is given in the next section. 

\subsubsection{Damage Detection} \label{neumann:sec_cond_mdKW51}
In this section, the Mahalanobis distance from Eq.~\ref{neumann:eq_mahalanobis} is applied for damage detection, combined with the temperature-dependent covariances from the previous section.
The modal frequencies of the first 200 recorded days before the retrofitting are used as the so-called Phase I, in-control data, and the Mahalanobis distance was calculated for each data point in this data set in a leave-one-out fashion. This means that for each sampling instance, the respective data point was excluded, and the remaining data points were used to estimate the (eight-dimensional) conditional mean vector and covariance matrix for the observed temperature. Although this procedure seems computationally expensive, it is a common approach for in-sample evaluation to avoid over-optimistic results. As mentioned above, we employed the bilinear approach for the conditional mean and the proposed non-parametric, kernel-based estimator for the conditional covariance. 
\begin{figure}[hbt]
    \centering
    \includegraphics[width=1\textwidth]{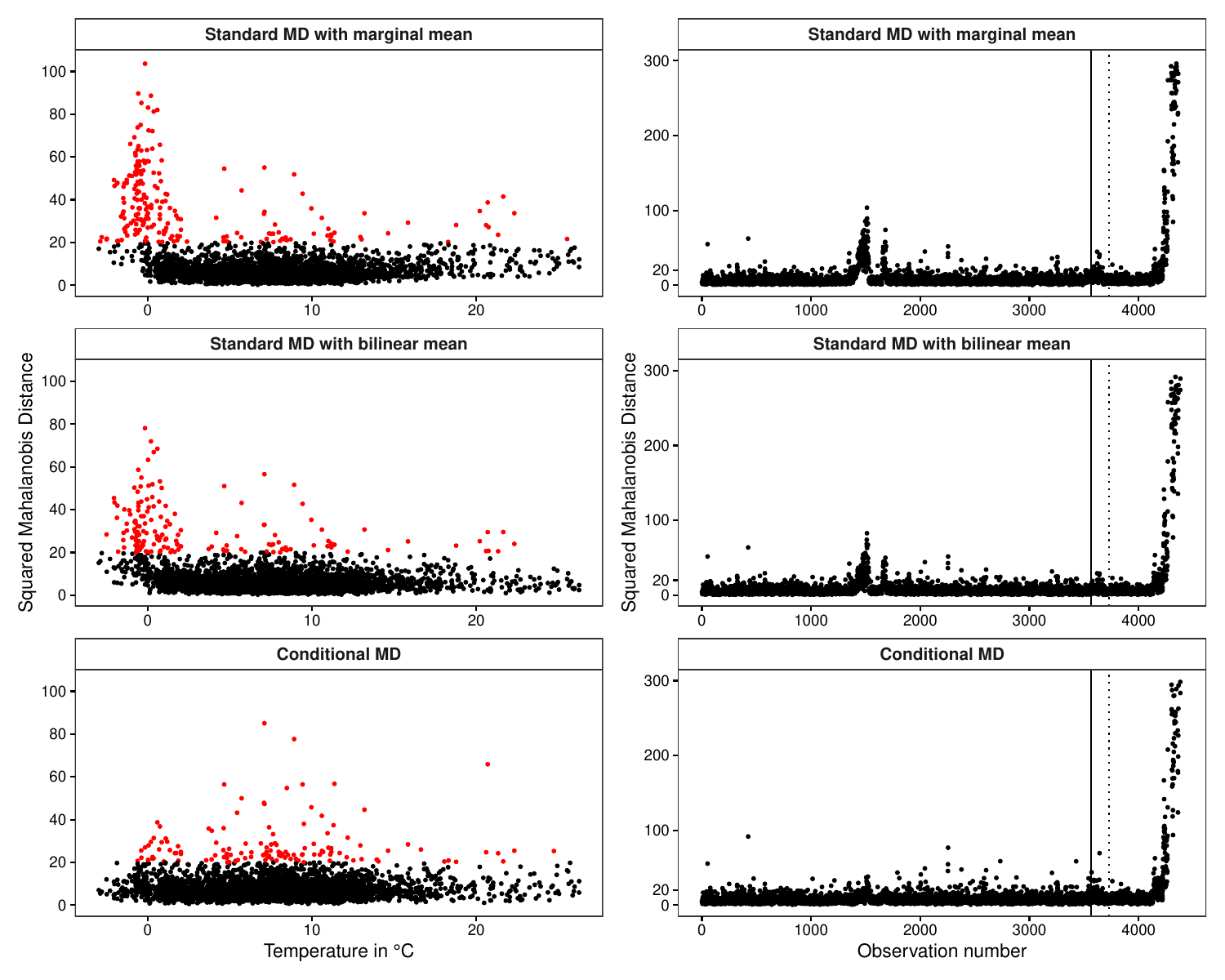}
    \caption{Comparing the Mahalanobis distance using the standard approach (top), the corrected mean value (center), and the corrected mean and covariance (bottom) on the KW51 data. The left column shows the results for Phase I, in-control data for varying temperature values (with false alarms according to the 99\%-quantile of the $\chi^2(8)$-distribution in red), while the right column shows the MD values over time, with the solid vertical line indicating the end of the in-control training phase; the dotted vertical line marks the beginning of the retrofitting, which eventually alters the structural properties of the bridge. For better visualization of the effects in the right column, (squared) MD values are cut at 300, and Phase II monitoring data is displayed until observation number 4400 only.}
    \label{neumann:fig14}
\end{figure}

The left side of Figure~\ref{neumann:fig14} displays the (squared) Mahalanobis distance against different temperature values before the retrofitting.
For illustration, diagnostic values above 20 are drawn in red. The value of 20 corresponds to the 99\%-quantile of the $\chi^2$-distribution with eight degrees of freedom (the theoretical null distribution for monitoring an eight-dimensional, normally distributed feature vector). The figure is split into three subplots, as three different versions of the Mahalanobis distance from Eq.~\eqref{neumann:eq_mahalanobis} are applied to highlight the advantages of the developed approach:
\begin{itemize}[noitemsep]
    \item[(a)] \textit{No Correction}: This is the standard version of the Mahalanobis distance using the marginal mean and the marginal covariance, see Figure~\ref{neumann:fig14} (top), which neglects temperature effects. 
    \item[(b)] \textit{Corrected Mean:} 
    The standard Mahalanobis distance is used in this scenario, but the temperature effect is removed using a bilinear response surface; see Figure~\ref{neumann:fig14} (center). 
    \item[(c)] \textit{Full Correction:} 
    In this scenario, the conditional Mahalanobis distance was applied, where both the mean and the covariance are corrected considering temperature effects, see Figure~\ref{neumann:fig14} (bottom).
\end{itemize}
As we can see (in Figure~\ref{neumann:fig14}, top), the standard Mahalanobis distance leads to extremely large MD values around $0^\circ$C, because the temperature has the most pronounced effects on modal frequencies for subzero temperatures. If the mean value is corrected, fewer values of the diagnostic exceed the limit of 20 on the in-control data, where no ``damage'' has occurred. However, the problem of having too many large test responses around $0^\circ$C is still present, see Figure~\ref{neumann:fig14} (center). A SHM system based on state-of-the-art temperature correction would misinterpret the large MD values as indicated damage, although these (false) alarms are caused by low temperatures. The conditional version of the Mahalanobis distance with both conditional mean and covariance leads to substantially fewer values beyond the limit of 20 for subzero temperatures, see Figure~\ref{neumann:fig14} (bottom). For temperatures below $2^\circ$C, for instance, the standard approach would lead to a false alarm rate of 38\%, compared to 25.8\% if the mean value is corrected, and 5.8\% for the conditional approach developed in this paper. For temperatures above $2^\circ$C, the false alarm rates are 1.6\%, 1.5\%, and 3.1\%, respectively, meaning the conditional MD leads to a larger number of false alarms. This is because the classical covariances are substantially larger compared to the conditional approach for warmer temperatures, which makes the test less sensitive.
For the entire Phase~I set combined, the false alarm rates are 6.3\%, 4.9\%, and 4\%, respectively. That means the developed approach yields comparable false alarm rates over the entire temperature range and substantially reduces the number of false alarms for cold temperatures compared to the standard approach. This is a desirable feature in bridge engineering, as the structure would not be closed for no reason, avoiding revenue losses. 

In theory, the number of (false) alarms should be 1\% because the threshold is set to be the 99\%-quantile of the $\chi^2$-distribution. This indicates that natural frequencies might still not be normally distributed but have more probability mass in the tails. This might be caused by further confounders beyond temperature, so-called ``unobserved heterogeneity''. Nevertheless, the conditional version of the Mahalanobis distance from Eq.~\eqref{neumann:eq_mahalanobis}, developed in this paper, best approximates the hypothetical null distribution and clearly outperforms state-of-the-art approaches, as the number of false alarms for temperatures below $2^\circ$C could be reduced by 75 to 85\% (23 false alarms instead of 103 or 152).

An important question is whether the developed approach only removes temperature-related information or whether damage-related information is removed as well. To analyze this, the right side of Figure~\ref{neumann:fig14} shows the sequential monitoring of the (squared) Mahalanobis distance over time for the three considered versions \mbox{(a)~--~(c)}. In SPC terminology, the squared Mahalanobis distance corresponds to a so-called Hotelling control chart \citep{Hotelling_1947}.
The vertical solid line in each chart indicates the end of the Phase I data set used so far and the beginning of the actual monitoring phase, also called Phase II. The mean and covariances estimated on the Phase I data are used to calculate the MD for Phase II data. 
The vertical dotted line marks the beginning of the retrofitting period. So, the period between the two vertical lines demonstrates how well the algorithm performs for previously unseen validation data from the undamaged, in-control state, while the data to the right of the dotted line displays the behavior of the Mahalanobis distance for test data with changing structural properties of the bridge. We can appreciate that the conditional approach of the Mahalanobis distance, Eq.~\eqref{neumann:eq_mahalanobis}, removes the temperature effect during the in-control phase but that the magnitude of the chart statistic remains the same for the testing data set with changing bridge conditions, meaning that ``damage'' during the retrofit is detected reliably and no damage-related information is removed. Specifically, if using the standard approach (with marginal or bilinear mean) about 6.7\% of the data points between the dashed vertical line in Figure~\ref{neumann:fig14} and observation number 4400 are above the threshold value of 20, while for the conditional approach, this number is even 8.8\%. This means that the conditional approach leads to a higher ``probability of detection'' here. At the same time, it is worth pointing out again that the temperature-induced artifacts (i.e., implausibly large MD values) around observation number 1,500 are removed using the new conditional approach.

\section{Conclusion}
\label{neumann:sec_conclusion}
The main contribution of this paper is a novel, non-parametric, kernel-based approach for identifying confounding effects in system response quantities. The approach requires the confounding variable to be measured (e.g., temperature, operational loads), and a distinguishing feature is that it captures changes in the means of the system response quantities and, more importantly, changes in the (co-)variances, using a conditional covariance matrix. The method requires one user-defined input parameter (vector), that is, the bandwidth(s) for the kernel, and an optimization scheme was outlined to give guidance on how to find appropriate values. 
Based on two well-established benchmark data sets from real bridges, it was demonstrated that the (co-)variances can change substantially for varying temperatures, in particular for subzero temperatures. This could be due to changes in the system response data or due to temperature-sensitive measurement equipment. Moreover, a numerical and easily reproducible case study was presented (a) to demonstrate that the underlying covariance function (the ground truth) can be retrieved correctly and (b) to validate the optimization scheme for the bandwidth. 

A second contribution is the further development of standard tools for anomaly detection, including the Mahalanobis distance and principal component analysis, such that they consider the temperature-dependent covariances of the system output. Typically, PCA is an unsupervised method that does not require the confounding variables to be measured. However, when combined with confounder-adjusted covariances, it turns into a supervised method that is able to exploit the additional information provided through measurements of the confounding variables. By applying existing approaches for the removal of confounding effects (e.g., response surface modeling) to the bridge case studies, it could be demonstrated that those methods may lead to a considerable number of false alarms--particularly for low temperatures in the case of the KW51 bridge. The newly developed method, on the other hand, reduced the risk of false alarms substantially for the KW51 bridge while retaining all damage-related information. Reducing false alarms is critical because each bridge closure due to false alarms leads to reduced acceptance of the monitoring procedure, discontent among users, and (depending on the toll system) a loss of revenue for operators. Furthermore, the developed method for confounder removal leads to system outputs that approximate normal distributions more closely, provided all confounding variables are measured. Often, data-driven algorithms for the removal of environmental effects are applied to raw system outputs, and subsequently, the extracted features are handed over to other damage diagnosis modules, e.g., for statistical process control. In many such cases, it is essential that the extracted features approximate Gaussian distributions. 
 
As shown in the paper, the developed method can be universally applied to static response measurements (strains, inclinations), dynamic response measurements (acceleration), or damage-sensitive features that are extracted from measurements (natural frequencies). This was demonstrated using data from the Munich Test Bridge and the KW51 railway bridge in Leuven. Future work particularly involves evaluating the new method's potential in detecting small damages/changes in the bridge's behavior (i.e., smaller than those present in the KW51 data)~\citep{Jaelani.etal_2024}. In this paper, we only considered the effect of one confounder variable (i.e., temperature) and disregarded other confounding effects. Extending the method to multiple confounders or temperature sensors will also be part of future research.

\section*{Software}
\label{neumann:sec_Software}
The data analysis was performed using the statistical software \texttt{R} \citep{R_2023} and add-on package \texttt{signal} \citep{Rsignal_2014}. 
\texttt{R} code implementing the estimate of the conditional covariance from Section~\ref{neumann:sec_nonoparm_est_cond_cov} together with an illustrative example reproducing the results from Figure~\ref{neumann:fig13} is available on GitHub at \url{https://github.com/neumannLizzie/conditional-covariance/}.

\section*{Data availability}\label
{neumann:sec_data_availability}
The eigenfrequency and environmental data for the railway bridge KW51 is available from \cite{Maes.Lombaert_2020}.
The acceleration, inclination, strain, and environmental data for the Test Bridge Munich is available at \url{https://github.com/imcs-compsim/munich-bridge-data}~\citep{Jaelani.etal_2023}.

\section*{Acknowledgements}
This research paper out of the project `SHM -- Digitalisierung und Überwachung von Infrastrukturbauwerken' is funded by dtec.bw -- Digitalization and Technology Research Center of the Bundeswehr, which we gratefully acknowledge. dtec.bw is funded by the European Union -- NextGenerationEU. We thank the hpc.bw team for their support within the joint dtec.bw subproject `HPC for semi-parametric statistical modeling on massive datasets'. The high-performance computing cluster HSUper has been established through the hpc.bw project, also funded by dtec.bw. We thank Dimitrios Anastasopoulos for providing the photo of the KW51 bridge.
We thank the two reviewers for their valuable comments, which have significantly improved the article.

\bibliographystyle{abbrvnat}
\bibliography{covest.bib}

\end{document}